\documentclass[11pt]{amsart}
\usepackage[letterpaper,margin=1in]{geometry}                
\usepackage{amssymb,amsmath,amsfonts}
\usepackage{xcolor}
\usepackage[colorlinks,
             linkcolor=blue,
             citecolor=green!70!black,
             pdfproducer={pdfLaTeX},
             pdfpagemode=None,
             bookmarksopen=true
             bookmarksnumbered=true]{hyperref}
\usepackage{graphicx}
\usepackage[utf8]{inputenc}

\newcommand\cP{\mathcal P}

\newcommand\bC{\mathbb C}
\newcommand\bH{\mathbb H}

\newcommand\bR{\mathbb R}
\newcommand\bZ{\mathbb Z}

\DeclareMathOperator\homology{H}
\renewcommand\H\homology

\DeclareMathOperator\cycles{Z}
\newcommand\Z\cycles

\DeclareMathOperator\chains{C}
\newcommand\C\chains

\renewcommand\d{\mathrm d}

\newcommand\mono\hookrightarrow
\newcommand\epi\twoheadrightarrow

\newcommand\<\langle
\renewcommand\>\rangle
\newcommand\sminus\smallsetminus

\DeclareMathOperator\Sq{Sq}

\newcommand\SV{\mathrm{SV}}
\newcommand\SA{\mathrm{SA}}

\DeclareMathOperator\Cliff{Cliff}
\newcommand\GP{\mathrm{GP}^\times}
\newcommand\bGP{\mathrm{bGP}^\times}
\newcommand\fGP{\mathrm{fGP}^\times}
\newcommand\Sre\GP

\DeclareMathOperator\id{id}

\newcommand\longto{\longrightarrow}

\newcommand\define[1]{\emph{#1}}

\usepackage{tikz}
\usetikzlibrary{decorations.markings,arrows,decorations.pathreplacing,arrows,decorations.pathmorphing,decorations.pathreplacing,decorations.markings,shapes.geometric,through,fit,shapes.symbols}
\tikzset{ribbon/.style={draw=white,double=purple,very thick,double distance=2pt}}
\tikzset{onearrowlate/.style={postaction={decorate}, decoration={markings,mark=at position .7 with {\arrow[draw,line width=1pt]{>}}}}}
\tikzset{onearrow/.style={postaction={decorate}, decoration={markings,mark=at position .55 with {\arrow[draw,line width=1pt]{>}}}}}

\makeatletter
\tikzset{
    fading speed/.code={
        \pgfmathtruncatemacro\tikz@startshading{50-(100-#1)*0.25}
        \pgfmathtruncatemacro\tikz@endshading{50+(100-#1)*0.25}
        \pgfdeclareverticalshading[%
            tikz@axis@top,tikz@axis@middle,tikz@axis@bottom%
        ]{axis#1}{100bp}{%
            color(0bp)=(tikz@axis@bottom);
            color(\tikz@startshading)=(tikz@axis@bottom);
            color(50bp)=(tikz@axis@middle);
            color(\tikz@endshading)=(tikz@axis@top);
            color(100bp)=(tikz@axis@top)
        }
        \tikzset{shading=axis#1}
    }
}
\makeatother

\newcommand{\be}{\begin{equation*}}
\newcommand{\ee}{\end{equation*}}
\newcommand{\ket}[1]{|#1\rangle}

\title{Symmetry Protected Topological phases and Generalized Cohomology}
\author{Davide Gaiotto and Theo Johnson-Freyd}
\date{\today}

\begin{document}
\begin{abstract}
We discuss the classification of SPT phases in condensed matter systems. 
We review Kitaev's argument that SPT phases are classified by a generalized cohomology theory,
valued in the spectrum of gapped physical systems \cite{KitaevTalk,KitaevTalk2}. We propose a concrete description of that spectrum 
and of the corresponding cohomology theory. We compare our proposal to pre-existing constructions in the literature.
\end{abstract}
\maketitle

\section{Introduction and Conclusions}

The general classification of Symmetry Protected Topological phases of matter is an important problem in theoretical
physics and mathematics \cite{SPT4,Kitaev:2009mg,SPT1,SPT2,Chen:2010gda,Chen:2011pg,SPT5,Chen:2011bcp,SPT7,Vishwanath:2012tq,Gu:2012ib,Wen:2013oza,SPT3,Wen:2013ue,Hung:2013cda,Chen:2013jha,SPT6,KitaevTalk,Kapustin:2014tfa,Kapustin:2014dxa,FH,Xiong:2017ab}. The typical approach to the problem in the condensed matter physics literature is constructive: produce explicit lattice Hamiltonians or lattice statistical mechanics models for the phases; identify observables which can demonstrate that the candidate phases are distinct from each other. This concrete approach has produced results of notable mathematical sophistication, with beautiful and robust structures which emerge unexpectedly from intricate calculations in ad hoc constructions. 

In a beautiful talk \cite{KitaevTalk}, Kitaev proposed an homotopy-theoretic classification of SPT phases, 
based on the notion of abstract {\it spectra} of invertible physical systems. An interesting aspect of the proposal is 
that the classification problem for SPT phases can be split into an ``easy'' part and a ``hard'' part. 
The ``hard'' part is to understand in sufficient detail invertible physical systems which 
have no internal symmetries which act on bosonic observables. The ``easy'' part is to add symmetry to the problem.

In the context of topological field theory, 
homotopy-theoretic ideas also lead to the classification of invertible phases in terms of cobordism spectra \cite{Kapustin:2014tfa,Kapustin:2014dxa,FH}, which can be understood in terms of ``topological effective actions'' or directly in terms of the axioms of topological field theory \cite{Seg,At}. 
The comparison between the two approaches raises a lot of difficult and interesting 
questions associated to the relation between gapped phases of matter, local quantum field theories, and abstract 
topological field theory. 
These cobordism proposals are not constructive:
 it is not known how to convert the homotopy theory definition 
of invertible TFT into a construction of phases of matter. Indeed, even if we consider  
a high energy physics version of the problem and focus on phases of gapped local quantum field theories, 
we would be still be unable to convert the homotopy theory definition into a definition of local QFT effective actions.
More pessimistically, some invertible homotopy TFTs may simply not be realizable as local lattice 
systems or even as local quantum field theories.

In this paper we will {\it not} make the assumption that gapped phases of matter can be fully captured by the language of 
topological field theory, although we do confirm in Section~\ref{sec:five} that certain classifications of gapped phases and of topological field theories agree {\it a posteriori}. Instead, we will extend Kitaev's proposal \cite{KitaevTalk,KitaevTalk2}, borrowing some ideas from TFT and high energy physics 
in order to provide a more concrete description of the spectra of invertible phases of matter. 
Our formulation of Kitaev's proposal is the following \footnote{See \cite{Xiong:2017ab} for an earlier discussion of Kitaev's proposal and its consequences.}: 

\begin{itemize}
\item STP phases in $d+1$ spacetime dimensions are classified by a reduced generalized cohomology group $\widetilde\H^{d+1}(B G_b; \Sre_{\leq d+1})$,
there $G_b$ is the bosonic symmetry of the theory, i.e.\ the quotient of the on-site internal symmetry group by fermion parity
operation.
\item The target $\Sre_{\leq d+1}$ of the generalized cohomology theory $\H^{\bullet}(-; \Sre_{\leq d+1})$ is essentially the {\it same} as the space of 
invertible phases of matter in spacetime-dimension $d+1$ or lower.  More precisely, 
these spaces combine into a {\it spectrum} $\Sre$ 
of invertible phases of matter in arbitrary dimensions, and $\widetilde\H^{d+1}(B G_b; \Sre_{\leq d+1}) = \widetilde\H^{d+1}(B G_b; \Sre)$ for degree reasons.
\end{itemize}

Our strategy is to fully implement the idea that SPT phases can constructed by ``decorating'' by invertible phases 
the domain walls for the $G_b$ internal symmetry \cite{CLV} and their junctions of various codimension. 
That perspective allows us to ``retract'' the space of invertible physical systems to a simpler model 
which only keeps track of phases of invertible defects of various dimensionality. 
From this perspective, the generalized cohomology theory above packages in a convenient fashion 
all the information needed to make sure the ``decoration'' of domain walls is self-consistent. 
In particular, we get a concrete simplicial description of the cohomology theory which is instrumental 
in making contact with previous constructions in the literature. 

A side effect of our calculations is that they demystify the surprise appearance of stable homotopy operations in the 
construction of fermionic SPT phases. In a seminal paper \cite{Gu:2012ib}, 
Gu and Wen gave a detailed construction of a class of fermionic SPT phases with unitary symmetry group. 
We will show in \S\ref{supercohomology section} that the Gu-Wen classification has a neat interpretation as $\H^{d+1}(B G_b; \Sre_{\leq 1})$: 
 it only takes into account the existence of  non-trivial fermionic invertible phases in $0+1$ dimensions. The final result of intricate computations in dimension up to $(3+1)d$ was a {\it supercohomology} theory, with a differential involving an exotic operation called the 
``Steenrod square''. We will see that the Steenrod square emerges naturally in any dimension as the only ``stable cohomology operation'' which could appear in the generalized cohomology theory. 

Subsequent constructions in \cite{SPT9,SPT10,SPT8,SPT11} take into account the existence of the Majorana chain in $1+1$ dimensions. The computation of the final cohomology theory is very intricate and dimension-specific and involves both the Steenrod square and an extra new operation. We identify it with $\H^{d+1}(B G_b; \Sre_{\leq 2})$ in any dimension.

In general, our analysis also allows one to compare classifications of SPT phases, invertible local QFTs and 
invertible homotopy TFTs by comparing the corresponding invertible
systems and invertible defects. 
Our discussion includes a variety of situations, including anti-unitary symmetries, symmetry groups which mix with 
fermion number symmetry, higher form symmetries, etc., simply by allowing $G_b$ to ``act'' non-trivially on the 
spectrum $\Sre$ \cite{KitaevTalk} or by replacing the classifying space $B G_b$ with other spaces \cite{Kapustin:2013uxa}. 
Part of our analysis applies to the more general problem of endowing some topological phase $P$ with the structure of a symmetry enriched phase for a symmetry $G_b$. The space of physical system belonging to the phase $P$ 
can be modelled up to homotopy as some topological space, which we can denote as $\mathrm{P}$, 
built from invertible defects in $P$. Then 
the $G_b$ symmetry enrichments of $P$ are classified by (homotopy classes of) maps from $BG_b$ to $\mathrm{P}$.
When $P$ is a trivial or invertible phase, then the calculation reduces to the calculation of a generalized cohomology theory.

\subsection{Outline of the paper}
In Section \ref{sec:two} we review the definitions of invertible phases of matter, topological defects and invertible topological defects. 
In Section \ref{sec:three} we review spectra and define the spectrum of invertible phases. 
In Section \ref{sec:four} we derive the classification of SPT phases as cohomology valued in the spectrum of invertible phases. 
Sections \ref{sec:five} and \ref{sec:six} are devoted to examples and to the computation of $\GP_{\leq n}$ for low $n$.  Section \ref{sec:anomalies} discusses the use of SPT phases as anomaly theories. 

\section{Generalities} \label{sec:two}

\subsection{Stacking and invertibility}
\begin{figure}[t]
\includegraphics[width=12cm]{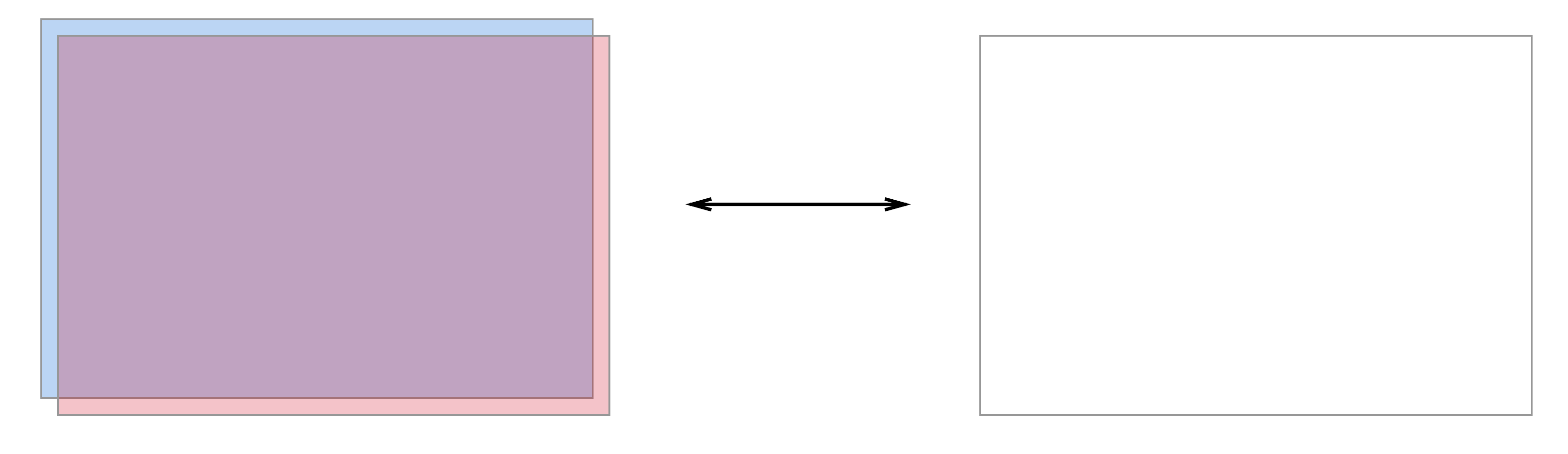}
\caption{A system (red) defines an invertible phase of matter if we can find another system (blue) such that the combination of the two system is in a trivial phase
}\label{fig:one}
\end{figure}

Recall some basic definitions:
\begin{itemize}
\item Two gapped systems are in the same phase if they can be continuously deformed into each other. 
\item A gapped system is considered to be in a trivial phase if it can be continuously deformed to a trivial gapped system, such as a system whose Hamiltonian is a sum of terms involving a single site of the lattice, with a single ground state and a large gap for each site. 
\item Any two gapped systems can be \define{stacked}, simply by taking the Hilbert space to be the tensor product of the Hilbert spaces of the two systems and the Hamiltonian to be the sum of the Hamiltonians. The stacking operation is a commutative and associative operations on gapped systems and thus on phases of matter. 
\item A gapped system $A$ is invertible if we can find another gapped system $A^{-1}$ such that stacking $A$ and $A^{-1}$ results in a trivial phase. See Figure \ref{fig:one}.
Invertible phases of matter form an Abelian group under the stacking operation. 
\end{itemize}

Note that the precise definition of ``continuous deformation'' of a gapped system is somewhat subtle. A typical interpretation is a combination of two types of operations:
\begin{itemize}
\item Deformations along which the gap remains bounded from below and no first-order phase transitions occur 
\item The operation of stacking with a trivial system.
\end{itemize}

The definition of ``phase'' and the topological considerations we use here and later on in the paper require a basic ``axiom'':
the existence of a well-defined ``space'' of gapped systems, which can be given a topology as we just described. 
It is useful to point out that alternative quantum-information theoretic definitions of the space of gapped systems are also available, based on the
properties of their vacuum wavefunctions \cite{KitaevTalk}. It is not known if different definitions are fully equivalent, 
but the general conceptual framework we use should remain valid.

Notice the crucial terminology distinction between ``system'' and ``phase''. A \emph{phase} is an equivalence class of systems and 
a system is a concrete physical realization of a phase. 

It is important to notice that the notion of ``phase'' is extracting only the roughest topological 
information about the space of gapped systems: it focusses on the set of connected components of that space. 
As we proceed in our discussion, we will often encounter situations where other topological aspects of 
the space of gapped systems have a measurable effect. This is the underlying motivation for the homotopy-theoretic approach we will employ in the rest of the note.

Notice that we have not defined the notion of ``gapped system''. It would be very interesting to find a concise way to formalize the properties of gapped systems 
which we use in the rest of our discussion, so that they could be formalized into some set of mathematical axioms. The most crucial property we expect from gapped system 
is some sort of ``locality'' in space and time. 
Concretely, this means that any gapped system which is sufficiently close to being periodic in space and time-independent at the microscopic scale
will look like a homogeneous system at macroscopic scales. Furthermore, gapped modifications of the system which 
have finite extent in some directions will appear as localized ``defects'' of various co-dimension at macroscopic scales.
Moreover,
we expect to be always allowed to independently introduce local modifications of the system at locations separated by macroscopic scales. 
Finally, we  expect to be allowed ``mesoscopic'' constructions, where we take some gapped system with some microscopic correlation length $\ell$ and produce a new system 
by adding some structure at a new mesoscopic scale much larger than $\ell$ but much smaller than the 
macroscopic scale at which we imagine studying the system.   

\subsection{SREs, invertible phases of matter, and invertible TFTs}

We should discuss a minor matter of terminology. Our operative definition of SPT phase 
is a gapped phase of matter which becomes trivial if we ignore the symmetry.
The phrase ``SRE'' is  used differently by different authors: \cite{Kapustin:2014tfa} uses ``SRE'' to mean \emph{invertible} phase\footnote{Kitaev's definition of ``SRE'' as meaning ``exactly one ground state''  is essentially equivalent to invertibility, by the main theorem of \cite{Schommer-Pries}.}, whereas for \cite{SPT1} ``SRE'' specifically means the trivial phase.
To avoid confusion, we will avoid the term ``SRE'' in favour of ``invertible phase.'' Furthermore, we will consider both situations where ignoring the symmetry 
makes the phase completely trivial and situations where ignoring the symmetry may still leave an invertible phase.

We will try to avoid conflating gapped phases of matter and topological field theories. 
Even for invertible phases and invertible TFTs, it is not obvious to us that every TFT should admit a realization as a condensed matter system.
This is especially true if we restrict ourselves to consider systems/theories which can be described by the (continuum limit of) 
a translation invariant ``lattice model,'' i.e.:
\begin{itemize}
\item A quantum mechanical system defined on a $d$-dimensional lattice with an Hamiltonian consisting
of a sum of local terms, i.e.\ operators acting on the degrees of freedom on a collection of lattice sites of bounded spatial extent or
\item A statistical mechanics system defined on a $(d+1)$-dimensional lattice with statistical weights which are the product of 
local terms, i.e.\ functions of the degrees of freedom on a collection of lattice sites of bounded spatial extent.
\end{itemize}

Conversely, it is not completely obvious that an invertible phase of matter should define an invertible TFT: a lattice model 
is defined naturally on a local patch of flat, non-relativistic, non-rotationally symmetric space-time and some extra structure is needed in order 
to place it on a general Euclidean space-time manifold. If we require the space-time manifold to be framed this extra structure is mostly automatic, 
but typical physically-meaningful topological field theories are isotropic, and the extra structure needed to place the theory on an oriented (say) manifold can be quite involved.

\subsection{Defects} 
In the following we will often employ the notion of local defect, defined as a modification of the system 
which is localized in the neighbourhood of some locus of non-zero 
codimension in space-time. A simple example of a defect may be a modification of the lattice Hamiltonian which affects all links which cross some given domain wall in the system. We will restrict ourselves to gapped defects which are locally translation invariant in the directions parallel to the defect. 

Gapped defects of a system can be organized into equivalence classes, analogous to the phases of bulk systems: 
two defects in a given system are equivalent if they can be continuously deformed into each other without closing the gap.
We will often be a bit sloppy and say ``defect'' when really mean ``equivalence class of defects'' or ``phase of defects.'' 
We will make the distinction sharp whenever it matters. 

We will be particularly interested in defects which are (framed) topological, in the sense that the physics below the gap is unaffected 
by mild topological manipulations of the defect locus, such as translations in a direction perpendicular to the defect 
or rotations by some small angle.

\begin{figure}[t]
\includegraphics[width=12cm]{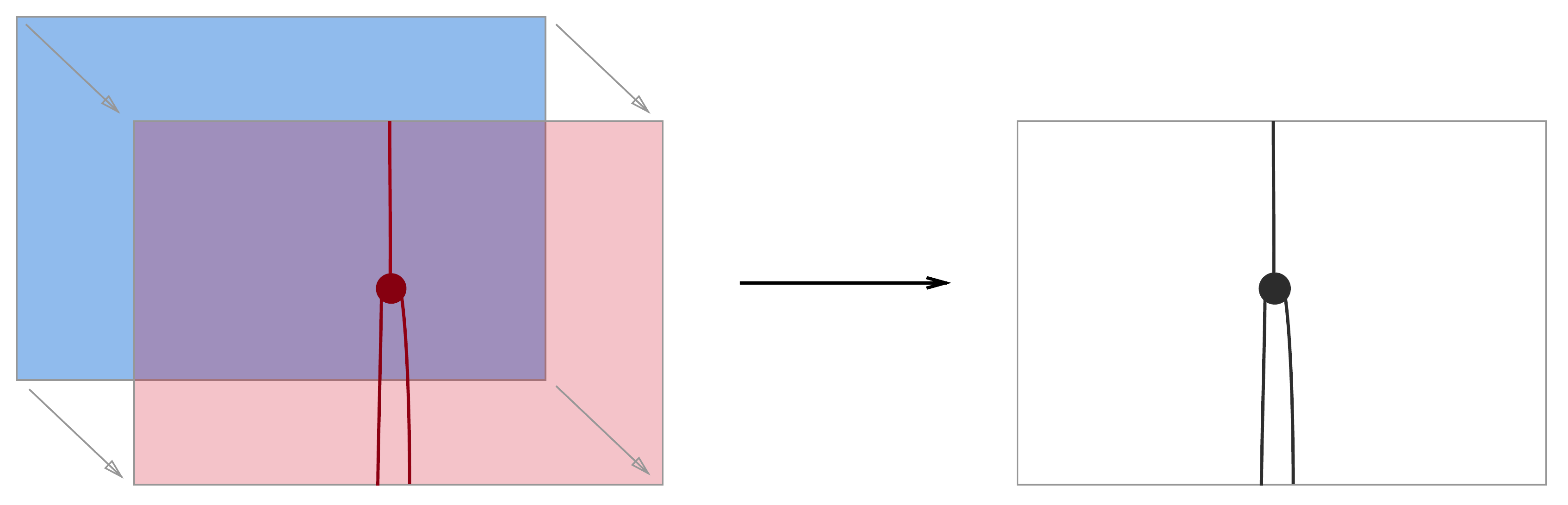}
\includegraphics[width=12cm]{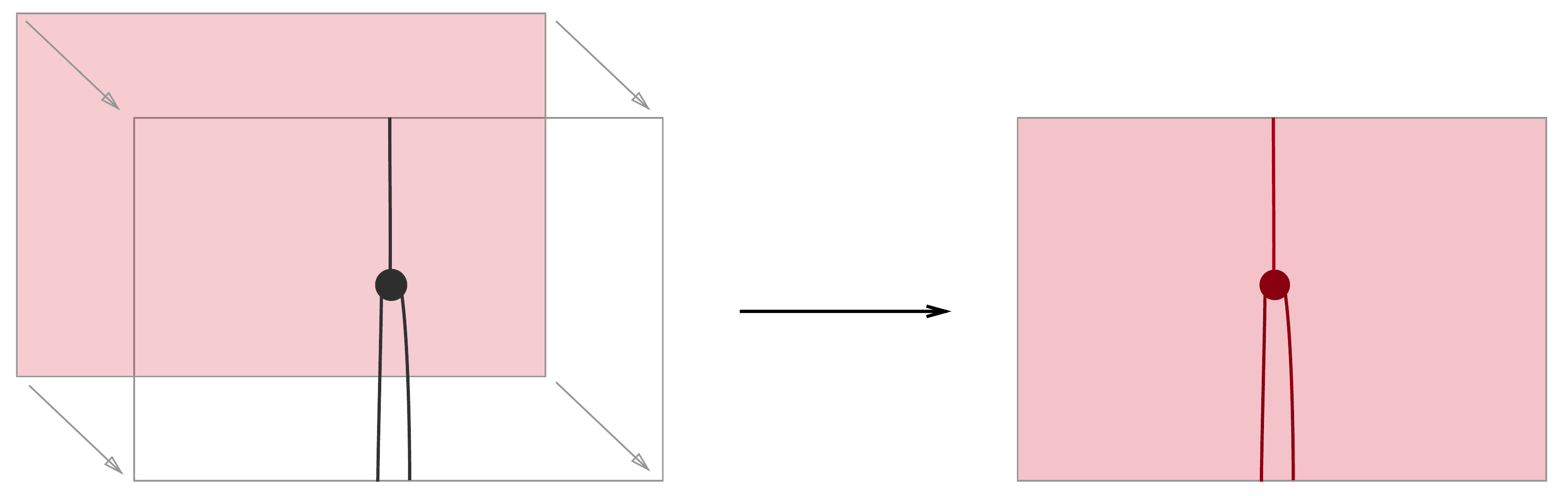}
\caption{Top: A collection of defects of various codimension in an invertible system can be mapped to a lower dimensional stand-alone 
system by stacking with an inverse system and deforming the bulk to a trivial system. Bottom: Lower dimensional systems can be stacked on top of a system to create simple defects. 
The two operations are essentially inverse of each other, up to important ambiguities in the choice of how to deform the bulk to a trivial system.
}\label{fig:two}
\end{figure}

These topological requirements can be reasonably expected to be true for generic gapped defects, 
but are essentially automatic for gapped defects in an invertible theory. Indeed, by stacking the whole system with the inverse of the original bulk theory 
we can map a defect of the invertible theory into a defect in the trivial theory and vice-versa. 
Furthermore, defects in a trivial theory are essentially the same as stand-alone lower-dimensional systems and are 
obviously (framed) topological. See Figure \ref{fig:two}.

This one-to-one map between defects in an invertible theory and lower-dimensional physical systems will be important to us. 
The reader should keep in mind that any concrete definition of this map will involve choices of how to deform systems into each other and that these choices will matter. 

It is also useful to consider defects-of-defects, such as lower-dimensional junctions between defects. An important consequence of topological invariance of defects under small rotations is that it allows us to think about junction between multiple defects and compositions of junctions, and junctions between junctions, etcetera.

\subsection{Invertible defects} \label{defects}
\begin{figure}[t]
\includegraphics[width=12cm]{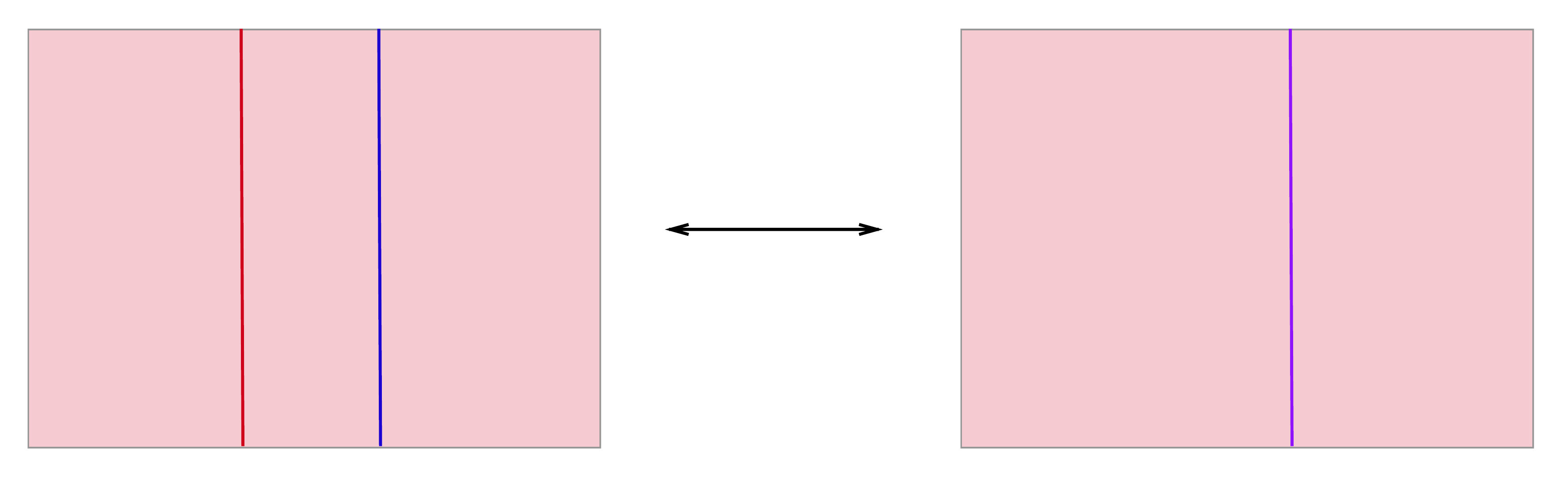}
\caption{Composition of defects}\label{fig:three}
\end{figure}

An important consequence of topological invariance under translations is that we can meaningfully ``compose'' defects by bringing them together. See Figure \ref{fig:three}.
Composition of defects is analogous to stacking of phases, except that it may in general be non-commutative. Things are much simpler 
for defects in invertible phases, as composition can be mapped to the stacking of the corresponding lower-dimensional phases. 

The notion of composition of defects leads to the notion of invertible defect: a defect is invertible if we can find an inverse defect, such that the composition of the two (in either order) can be continuously deformed to the trivial defect, i.e.\ to the system without a defect. 

A particularly important class of defects are codimension $1$ interfaces between different systems. The existence of an invertible interface between two systems implies a very strong relation between the two systems. Indeed, as explained also by~\cite{KitaevTalk}, we claim that it is equivalent to the two systems being in the same phase!%
\footnote{In TFT, the existence of an invertible interface between two systems is often taken to be the {\it definition} of equivalence between two theories.}
\begin{itemize}
\item Consider two systems $A$ and $B$ in the same phase. They must be connected by a continuous deformation. Pick a direction $x$ in space.
Consider a hybrid system which is identical to $A$ for $x\ll0$ and identical to $B$ for $x\gg0$, but transitions very slowly (i.e.\ over mesoscopic 
scales much larger than the correlation length of the system but much shorter than the scale at which we study the system)
through the family as $x$ moves from negative to positive. At sufficiently large scale, this setup defines a gapped interface between $A$ and $B$. The inverse 
interface is defined just in the same manner, following the family in the opposite direction. See Figure~\ref{fig:four}.
\item Consider two systems $A$ and $B$ related by an invertible interface $F$. Pick a direction $x$ in space. Consider a hybrid system consisting of 
a mesoscopic lattice of alternating slabs of $A$ and $B$ joined by $F$ and $F^{-1}$ interfaces. This hybrid system can be continuously deformed into $A$ by composing each $F$ 
interface with the inverse interface to its left. It can also be continuously deformed into $B$ by composing each $F$ 
interface with the inverse interface to its right. This provides a family of continuous deformations relating $A$ and $B$.
See Figure~\ref{fig:five}.
\end{itemize}
\begin{figure}
\begin{tikzpicture}
  \shade[left color=red!50, right color=blue!50, fading speed=50] (-6,0) rectangle (-2,3);
  \fill[color=red!50] (2,0) rectangle (4,3);   \fill[color=blue!50] (4,0) rectangle (6,3);
  \draw[thick, color=red!50!blue] (4,0) -- (4,3);
  \draw[{stealth}-stealth,thick] (-1,1.5) -- (1,1.5);
\end{tikzpicture}
\caption{A continuous interpolation between two systems can be identified with an invertible interface at a sufficiently large scale.}\label{fig:four}
\end{figure}
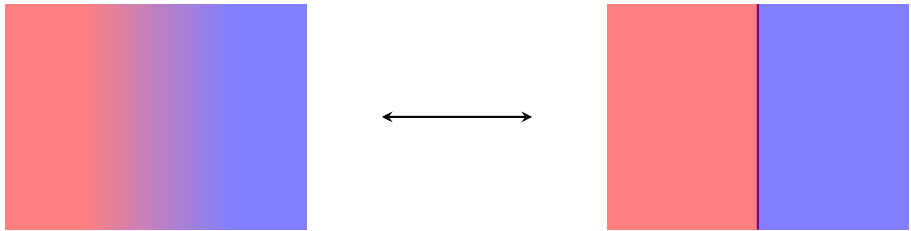
\begin{figure}[t]
\begin{tikzpicture}
  \foreach \x in {.1,.3,...,2.9}{
    \fill[color=red!50] (\x,0) rectangle +(.1,2);
    \fill[color=blue!50] (\x,0) rectangle +(-.1,2);
    \draw[thick, color=red!50!blue] (\x,0) -- +(0,2);
  }
  \foreach \x in {.2,.4,...,2.8}\draw[thick, color=red!50!blue] (\x,0) -- +(0,2);
  \fill[color=red!50] (-5,0) rectangle +(3,2);
  \fill[color=blue!50] (5,0) rectangle +(3,2);
  \draw[{stealth}-stealth,thick] (-1.5,1) -- +(1,0);
  \draw[{stealth}-stealth,thick] (3.5,1) -- +(1,0);
\end{tikzpicture}
\caption{A mesoscopic lattice of invertible interfaces between two systems realizes a continuous interpolation between them.}\label{fig:five}
\end{figure}
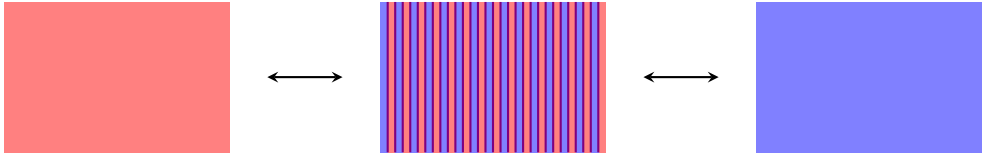

In particular, invertible interfaces allow us a glimpse into the non-trivial topology of 
the space of gapped systems: the set of equivalence classed of invertible interfaces between two systems 
gives us a handle on the space of inequivalent deformation paths relating the two systems. 

Similar considerations apply to invertible junctions between interfaces, etcetera:
phases of invertible defects of various dimensions in a given phase of matter encode the topology (or better, homotopy) of the corresponding 
space of gapped systems.

If we restrict ourselves to invertible systems, we have an interesting iterative structure: 
\begin{itemize}
\item Invertible interfaces between $d$-dimensional invertible systems exist only if the systems are in the same phase. Such interfaces 
can be put in correspondence with $(d-1)$-dimensional invertible systems, by the ``stacking with the inverse phase'' trick. 
\item Invertible junctions of invertible interfaces between $d$-dimensional invertible systems exist only if the 
interfaces are equivalent. They are in correspondence with $(d-2)$-dimensional invertible systems.
\item Etcetera.
\end{itemize}

We will now discuss the topology of the space of invertible gapped systems and the conjectural encoding 
in terms of invertible phases of systems and defects. 

\section{The spectrum of invertible phases} \label{sec:three}

\subsection{Definition of the spectrum} \label{definition of spectra}
A \define{spectrum}%
\footnote{There are many models of spectra; the one we are describing is called ``$\Omega$-spectrum.''}
$T$ in algebraic topology consists of a sequence of topological spaces $\dots, T_{-1}, T_0, T_1, T_2, \dots$ together with the following data.
First, each space $T_n$ should be equipped with a distinguished basepoint $0 \in T_n$. Second, let $\Omega T_n$ denote the space of loops in $T_n$ that start and end at $0$; then we should equip the sequence $T_\bullet$ with homotopy equivalences $T_{n-1} \overset\sim\to \Omega T_n$. In particular, the space $T_n$ determines all $T_m$ for $m < n$ up to homotopy equivalence, and the $T_m$ for $m > n$ witness $T_n$ as an infinite loop space (the homotopy version of ``topological abelian group'').
For any space $X$ and any $k \geq 0$, $\pi_k \Omega X = \pi_{k+1} X$. This allows one to define both positive and negative \define{homotopy groups} of $T$ by $\pi_k T := \pi_0 T_{-k}$. A spectrum is \define{connective} if all of its negative homotopy groups vanish, in which case it consists of the same data as the infinite loop space $T_0$. 
The \define{suspension} $\Sigma T$ of a spectrum $T$ is defined by $(\Sigma T)_n = T_{n+1}$.

We now describe the spectrum $\GP$ of invertible gapped phases of matter. Let $\GP_n$ denote the space of $n$-spacetime-dimensional invertible gapped systems. 
As mentioned in the previous section, continuous deformations within $\GP_n$ may include stacking with trivial systems or more generally operations which add and remove states 
at energies well above the gap. The basepoint of the space $\GP_n$ is the trivial system. 

As we discussed in the previous section, there is an obvious map $\Omega\GP_n \to \GP_{n-1}$. First choose a direction $x$ in space. Now, an element of $\Omega\GP_n$ is a continuous family of states of matter that begins and ends at the trivial state. Given such a family, consider a system on which for $x \ll 0$ and $x \gg 0$ is in the trivial state, but which transitions through the family as $x$ moves from negative to positive over mesoscopic scales much larger than the microscopic scale. 
At macroscopic length scales, this continuous transition can be squeezed into an $(n-1)$-spacetime-dimensional system. The map $\Omega\GP_n \to \GP_{n-1}$ is pictured in Figure~\ref{fig:four}, where we take both $A$ and $B$ to be the trivial system, but use a nontrivial deformation.

We claim that this map $\Omega\GP_n \to \GP_{n-1}$ is a homotopy equivalence. It suffices to give a claimed homotopy inverse $\GP_{n-1} \to \Omega\GP_n$ and show that both compositions $\Omega\GP_n \to \GP_{n-1} \to \Omega\GP_n$ and $\GP_{n-1} \to \Omega\GP_n \to \GP_{n-1}$ are homotopic to the identity. 

The map $\GP_{n-1} \to \Omega\GP_n$ is described in Figure~\ref{fig:five}: given an invertible $n-1$ dimensional phase $F$, treat it as an invertible defect in the trivial $n$-dimensional phase; interleave copies of $F$ and $F^{-1}$, separated at a mesoscopic scale; observe that this $F/F^{-1}$ system is connected to the trivial system by a continuous deformation in two different ways, and hence gives a loop in $\GP_n$ from the trivial system to itself. 

Finally, one can give explicit null homotopies for the compositions $\Omega\GP_n \to \GP_{n-1} \to \Omega\GP_n$ and $\GP_{n-1} \to \Omega\GP_n \to \GP_{n-1}$. We argue more generally that, given systems $A$ and $B$, the maps pictured in Figures~\ref{fig:four} and~\ref{fig:five} are homotopy inverses. We can sketch the basic arguments here:
\begin{itemize}
\item The composition $\{$paths from $A$ to $B\} \to \{$invertible defects between $A$ and $B\} \to \{$paths from $A$ to $B\}$, generalizing the composition $\Omega\GP_n \to \GP_{n-1} \to \Omega\GP_n$, turns a path $\cP$ into a new path constructed as follows. Choose a mesoscopic 1-dimensional lattice in the $x$-direction. Start at the system $A$, and then adiabatically transition to system $B$ \emph{just in the neighborhoods of the mesoscopic lattice points}. Now adiabatically expand those $B$-regions until they fill the system is purely in the $B$ state. This gives the new path from $A$ to $B$. It is canonically homotopic to the old path. See Figure \ref{fig:composition1}.
\begin{figure}[t]
\begin{tikzpicture}
  \shade[bottom color=red!50, top color=blue!50, fading speed=50] (-7,.5) rectangle (-2,3.5);
\draw[|-stealth,thick] (-1.5,2) -- +(1,0);
\clip (0,.5) rectangle (5,3.5);
   \fill[color = red!50] (0,0) -- (0,3) 
   cos ++(.25,-1) sin ++(.25,-1) cos ++(.25,1) sin ++(.25,1)
   cos ++(.25,-1) sin ++(.25,-1) cos ++(.25,1) sin ++(.25,1)
   cos ++(.25,-1) sin ++(.25,-1) cos ++(.25,1) sin ++(.25,1)
   cos ++(.25,-1) sin ++(.25,-1) cos ++(.25,1) sin ++(.25,1)
   cos ++(.25,-1) sin ++(.25,-1) cos ++(.25,1) sin ++(.25,1)
   -- (5,0) -- cycle
   ;
   \fill[color = blue!50] (0,4) -- (0,3) 
   cos ++(.25,-1) sin ++(.25,-1) cos ++(.25,1) sin ++(.25,1)
   cos ++(.25,-1) sin ++(.25,-1) cos ++(.25,1) sin ++(.25,1)
   cos ++(.25,-1) sin ++(.25,-1) cos ++(.25,1) sin ++(.25,1)
   cos ++(.25,-1) sin ++(.25,-1) cos ++(.25,1) sin ++(.25,1)
   cos ++(.25,-1) sin ++(.25,-1) cos ++(.25,1) sin ++(.25,1)
   -- (5,4) -- cycle
   ;
   \foreach \x in {0,.25,...,10}
   {
   \draw[color = red!50!blue!50, line width = \x, opacity=.05]
   (0,3) 
   cos ++(.25,-1) sin ++(.25,-1) cos ++(.25,1) sin ++(.25,1)
   cos ++(.25,-1) sin ++(.25,-1) cos ++(.25,1) sin ++(.25,1)
   cos ++(.25,-1) sin ++(.25,-1) cos ++(.25,1) sin ++(.25,1)
   cos ++(.25,-1) sin ++(.25,-1) cos ++(.25,1) sin ++(.25,1)
   cos ++(.25,-1) sin ++(.25,-1) cos ++(.25,1) sin ++(.25,1)
   -- (5,3)
   ;
   }
\end{tikzpicture}
\caption{The new path produced by the composition $\{$paths from $A$ to $B\} \to \{$invertible defects between $A$ and $B\} \to \{$paths from $A$ to $B\}$ from an old path from system $A$ (red) to system $B$ (blue). The horizontal direction denotes the spatial direction $x$ used to define the map $\{$paths from $A$ to $B\} \to \{$invertible defects between $A$ and $B\}$, and the vertical direction denotes the ``internal time'' parameter along which the path from $A$ to $B$ transforms.}\label{fig:composition1}
\end{figure}
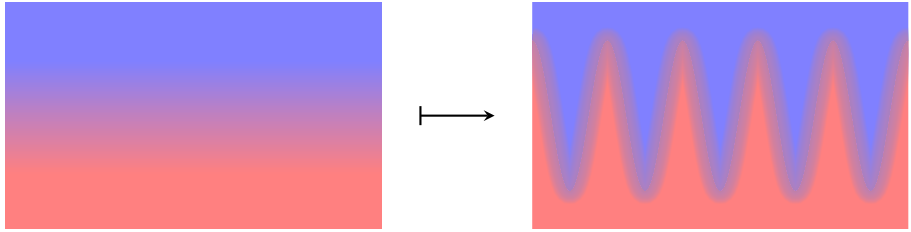
\item The composition $\{$invertible defects between $A$ and $B\} \to \{$paths from $A$ to $B\} \to \{$invertible defects between $A$ and $B\}$, generalizing the composition $\GP_{n-1} \to \Omega\GP_n \to \GP_{n-1}$, maps an invertible defect $X$ between $A$ and $B$ to a 
locally periodic mesoscopic sequence of interfaces $X$ and $X^{-1}$, beginning and ending with $X$, such that the ratio between the
$X-X^{-1}$ and $X^{-1} - X$ distances evolve from very small to very large from left to right; this mesoscopic sequence is then considered a single defect at large scales. The homotopy transforming this new defect into the old one is given by canceling interfaces $X$ and $X^{-1}$ where we keep the middle $X$ interface fixed and evolve the ratio to be very small 
uniformly to the left of it, very large uniformly to the right of it.  See Figure \ref{fig:composition2}.
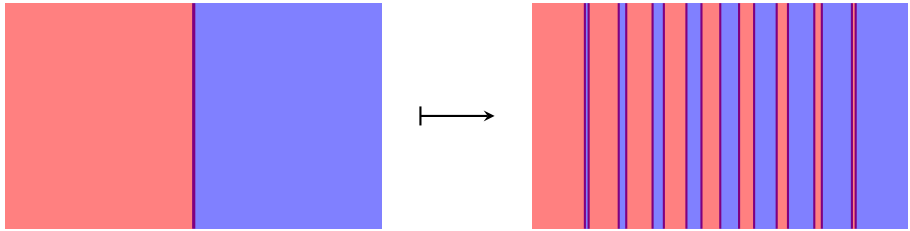
\begin{figure}[t]
\begin{tikzpicture}
  \fill[color=red!50] (-7,0) rectangle +(2.5,3);
  \fill[color=blue!50] (-4.5,0) rectangle +(2.5,3);
  \draw[very thick, color=red!50!blue] (-4.5,0) -- +(0,3);
  \draw[|-stealth,thick] (-1.5,1.5) -- +(1,0);
  \fill[color=red!50] (0,0) rectangle (.7,3);
  \fill[color=red!50] (.75,0) rectangle (1.15,3);
  \fill[color=red!50] (1.25,0) rectangle (1.6,3);
  \fill[color=red!50] (1.75,0) rectangle (2.05,3);
  \fill[color=red!50] (2.25,0) rectangle (2.5,3);
  \fill[color=red!50] (2.75,0) rectangle (2.95,3);
  \fill[color=red!50] (3.25,0) rectangle (3.4,3);
  \fill[color=red!50] (3.75,0) rectangle (3.85,3);
  \fill[color=red!50] (4.25,0) rectangle (4.3,3);
  \fill[color=blue!50] (.7,0) rectangle (.75,3);
  \fill[color=blue!50] (1.15,0) rectangle (1.25,3);
  \fill[color=blue!50] (1.6,0) rectangle (1.75,3);
  \fill[color=blue!50] (2.05,0) rectangle (2.25,3);
  \fill[color=blue!50] (2.5,0) rectangle (2.75,3);
  \fill[color=blue!50] (2.95,0) rectangle (3.25,3);
  \fill[color=blue!50] (3.4,0) rectangle (3.75,3);
  \fill[color=blue!50] (3.85,0) rectangle (4.25,3);
  \fill[color=blue!50] (4.3,0) rectangle (5,3);
  \foreach \x in {.7, .75, 1.15, 1.25, 1.6, 1.75, 2.05, 2.25, 2.5, 2.75, 2.95, 3.25, 3.4, 3.75, 3.85, 4.25, 4.3}{
  \draw[thick, color=red!50!blue] (\x,0) -- +(0,3);
  }
\end{tikzpicture}
\caption{The output of the composition $\{$invertible defects between $A$ and $B\} \to \{$paths from $A$ to $B\} \to \{$invertible defects between $A$ and $B\}$.}\label{fig:composition2}
\end{figure}
\end{itemize}

Since $\Omega\GP_n \to \GP_{n-1}$ is a homotopy equivalence, we achieve a spectrum $\GP$. By construction, its homotopy groups are $\pi_{-k} \GP = \pi_0 \GP_k = $ set of $k$-dimensional phases of matter, with the abelian group structure on $\pi_{-k}\GP$ given by stacking. 

The equivalence $\Omega\GP_n \to \GP_{n-1}$ implies that, as an infinite loop space, $\GP_n$ has another description. Namely, consider the higher category whose objects are $n$-dimensional invertible phases of matter, 1-morphisms are phases of $(n-1)$-dimensional invertible defects, 2-morphisms are phases of defects between defects, etc., together with its symmetric monoidal structure given by stacking and composition of defects. Since all objects and morphisms are invertible, this defines  a symmetric monoidal higher groupoid, or equivalently a connective spectrum. Since this category is built out of gapped phases of dimensions $0,\dots,n$, we will call it $\GP_{\leq n}$.

Actually, it is most convenient to index the corresponding spectrum so that the $n$-morphisms in the category contribute to degree-$0$ homotopy: we will use the symbol $\GP_{\leq n}$ to denote the spectrum formed by taking the infinite loop space corresponding to the category of $(\leq n)$-dimensional phases of matter and desuspending it. By construction, $\pi_k \GP_{\leq n} = \pi_k \GP = \pi_0 \GP_k$ if $k \geq -n$ and $\pi_k \GP_{\leq n} = 0$ if $k < -n$. The passage $\GP \leadsto \GP_{\leq n}$ is an example of taking a \define{connective cover}: if $T$ is a spectrum, then its $m$th \define{connective cover} is the unique spectrum $T \langle m \rangle$ equipped with a map $T \langle m \rangle \to T$ such that $\pi_k T\langle m \rangle = \pi_k T$ for $k \geq m$ and $\pi_k T \langle m \rangle = 0$ for $k < m$.

We would like to stress that the description of $\GP_{\leq n}$ as a groupoid of phases of invertible defects 
is eminently more manageable than an abstract description as a ``space'' of gapped systems. The space of gapped systems
is enormous and potentially very intricate. The collection of phases of invertible theories and invertible defects 
up to some fixed dimension is a discrete and manageable object. 

At any time, we can compile an approximation of $\GP_{\leq n}$ out of known phases and defects. As our knowledge of 
phases of matter improves, we can update our model for $\GP_{\leq n}$ accordingly. 

\subsection{Cohomology with coefficients in $\GP$ and internal symmetries in lattice models}

The main purpose for spectra in algebraic topology is to provide the coefficients for generalized cohomology theories. If $X$ is a topological space and $T = \{\dots,T_0,T_1,\dots\}$ is a spectrum, then the spaces $[X,T_n]$ of continuous maps $X \to T_n$ form a new spectrum $[X,T]$, and the \define{cohomology} of $X$ with coefficients in $T$ is by definition $\H^k(X;T) = \pi_k [X,T] = \pi_{k+n}[X,T_n]$ for $n \gg 0$. Let us focus on the spectrum $\GP$ of invertible phases of matter. Suppose that we describe $X$ by some cell complex. We claim that $\H^\bullet(X;\GP)$ describes phases coupled to a background field valued in $X$. Indeed, for each point (0-cell) in $X$ we can ask: ``how does our theory look like 
if the background field maps
all space-time to a neighbourhood of that point?'' The answers will be some invertible theories, one for each point.
For each segment (1-cell) between two points  in $X$ we can ask: ``how does our theory look like 
if half of space-time is mapped to a neighbourhood of one point, the other half to a neighbourhood of the other point, and the interface
region to a neighbourhood of the 1-cell?'' The answers will be some  invertible defects, one for each 1-cell.
Etcetera.

Notice that this claim is constructive. On one hand, given an invertible gapped system which can be coupled to maps from space-time into 
$X$, we can read off the data defining an element of $\H^\bullet(X;\GP)$, simply by coupling the system to appropriate maps and reading off the 
corresponding low energy phases. On the other hand, given an element in $\H^\bullet(X;\GP)$ and a cell-complex definition of $X$,
we can pick a representative element in $[X,\GP_n]$ and {\it build} a gapped system which reproduces it, with the usual trick of building a mesoscopic lattice system. For example, in a statistical mechanics 
setup:
\begin{itemize}
\item On facets of maximal dimension $D$ in the mesoscopic lattice, we put degrees of freedom valued in 0-cells of $X$.
\item On facets of dimension $D-1$ in the mesoscopic lattice, we put degrees of freedom valued in 1-cells of $X$.
\item Etcetera: the degrees of freedom on the mesoscopic lattice are the data of a map from the dual lattice to the cell complex for $X$.
\item We couple the mesoscopic lattice to a microscopic Hamiltonian in such a way that the microscopic theory in each 
$D$-dimensional facet is in the invertible phase  determined by the map from $X$ to $\GP$, with interfaces at 
each $(D-1)$-dimensional facet determined in a corresponding way, etc. 
\end{itemize}

By sufficiently relaxing our notion of target space $X$ to include classifying spaces $BG$ for a group $G$, we can apply this discussion to SPT phases, which are systems with a $G$-symmetry; we will classify SPT phases in terms of the cohomology of $BG$ with coefficients in $\GP$
\cite{KitaevTalk}. 
To an algebraic topologist this is a tautology: for any type of mathematical object, an object of that type with $G$ symmetry ``is'' a map from $B G$ to the space of all objects of that type. 
In the next section we connect this tautology (in the case of invertible phases) with a much more explicit description of SPT phases in terms of a cochain model of $\H^\bullet(-;\GP)$.

\section{SPT phases and spectra} \label{sec:four}

In this section we will discuss a very simple observation with deep consequences: the symmetries of a gapped system can be fully encoded in 
a certain collection of invertible defects.

\subsection{Symmetries and defects} \label{section symmetries and defects}

The standard notion of a ``non-anomalous'' internal symmetry of a lattice model is a symmetry 
which acts independently on the local degrees of freedom at each lattice site. Such a symmetry can always be gauged, 
by adding group degrees of freedom on links of the lattice and adjusting the Hamiltonian or statistical weights accordingly.%
\footnote{We include symmetries which act anti-unitarily on the Hilbert space, such as time-reversal symmetry. A cautious reader may bristle 
at the idea of gauging an anti-unitary symmetry. How does one define a connection for time-reversal symmetry? For a flat connection, this is actually possible. 
It is useful to think about coupling a lattice system to a flat connection as a modification of the tensor product used to assemble the full Hilbert space from the 
on-site Hilbert spaces: we keep the Hamiltonian unchanged but postulate that operators act as 
\begin{equation*}
(O_1 \otimes O_2)  \circ (v_1 \otimes_g v_2) = (O_1 v_1) \otimes_g (g^{-1} O_2 g v_2)
\end{equation*}
For anti-unitary symmetries, we can use a tensor product which is linear in one entry and anti-linear in the second entry. 
Concretely, that means that in the presence of domain walls for anti-unitary symmetries we will combine the Hilbert spaces for sites on the two sides of the walls
by a ``twisted tensor product'' which is anti-linear in the second entry. \label{foot}
}

\begin{figure}[t]
\includegraphics[width=16cm]{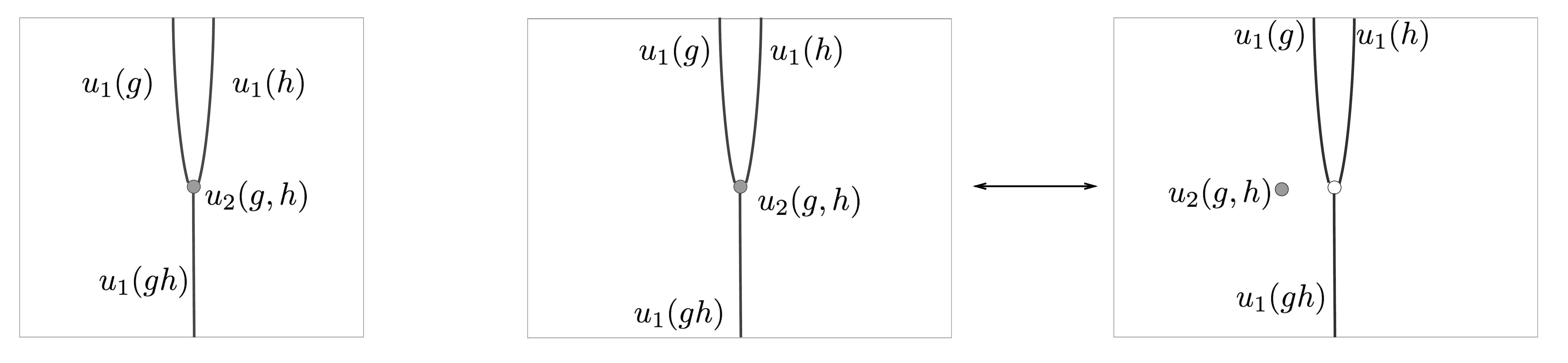}
\caption{Left: a localized flat connection gives rise to invertible domain walls and junctions labelled by group elements. Domain walls in the trivial phase are identified with systems $u_1(g)$ in one dimension lower. Right: Junctions between domain walls can be mapped to systems $u_2(g,h)$ in two dimensions lower at the cost of a non-canonical choice of reference junction (empty node) between the domain walls.}\label{fig:six}
\end{figure}

Whenever we have such an internal symmetry $G$, the system comes equipped naturally with a collection of invertible topological defect of various dimension,
 defined by turning on a background flat gauge connection localized in the neighbourhood of the defect (see Figure \ref{fig:six}): 
\begin{itemize}
\item Defects $u_1(g)$ of codimension $1$, labelled by a group element $g\in G$. They are defined by turning on a background connection with $g$ 
on all links which cross the defect.
\item Junctions $u_2(g_1,g_2)$ of codimension $2$ between defects $u_1(g_1)$, $u_1(g_2)$ and $u_1(g_1 g_2)$. They are defined by  turning on a 
background connection with $g_1$ on all links which cross $u_1(g_1)$, $g_2$ on all links which cross $u_1(g_2)$, $g_1 g_2$ on all links which cross $u_1(g_1 g_2)$.
\item Junctions $u_2(g_1,g_2,g_3)$ of codimension $3$ between the codimension $2$ junctions between defects $u_1(g_1)$, $u_1(g_2)$, $u_1(g_1 g_2)$,
$u_1(g_3)$, $u_1(g_2 g_3)$ and $u_1(g_1 g_2 g_3)$.
\item For general codimension $n$, junctions $u_n(g_1,g_2,g_3, \cdots, g_n)$ between codimension $n-1$ junctions between $\cdots$.
\end{itemize}

\begin{figure}[t]
\includegraphics[width=16cm]{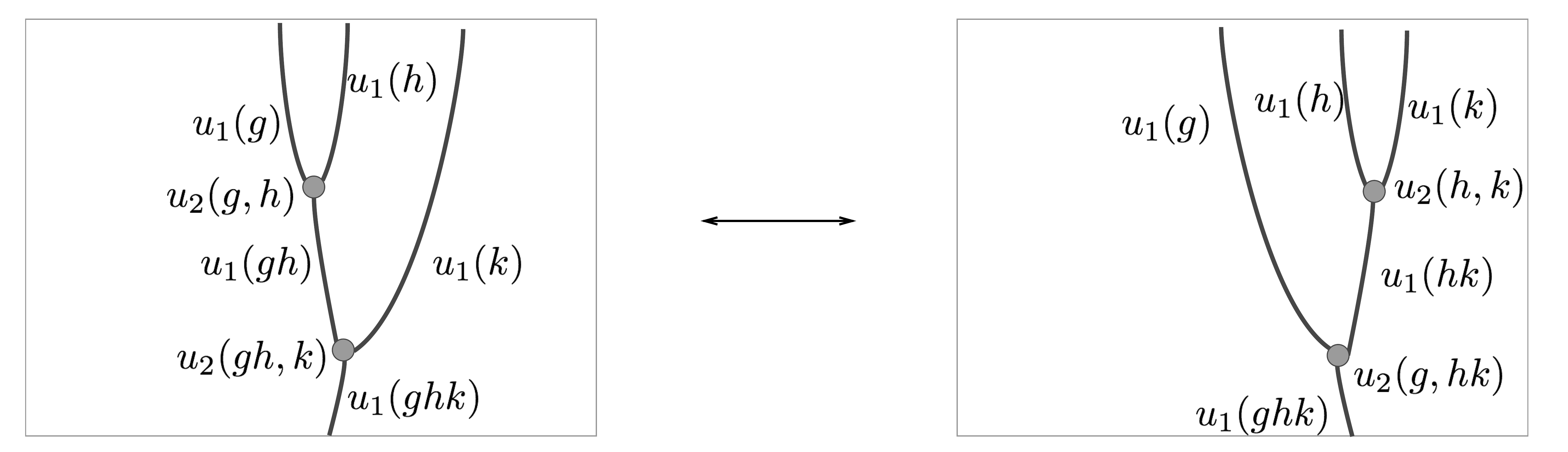}
\caption{Associativity for domain wall junctions}\label{fig:seven}
\end{figure}

All these junctions have an important property, which follows from gauge invariance of the underlying background connection: 
they are fully topological and any network of such junctions can be freely re-arranged (See Figure \ref{fig:seven}):
\begin{itemize}
\item Codimension $1$ defects compose according to the group law: $u_{g_1} \circ u_{g_2} \equiv u_{g_1 g_2}$ with $u_e$ being the trivial defect. 
The local equivalence of the defects is implemented by $u_2(g_1,g_2)$. 
\item Codimension $2$ defects associate: $u_2(g_1,g_2) \circ_{u_1(g_1 g_2)} u_2(g_1 g_2,g_3) \equiv u_2(g_2,g_3) \circ_{u_1(g_2 g_3)} u_2(g_1, g_2 g_3)$.
The local equivalence of the composite junctions is implemented by $u_3(g_1,g_2,g_3)$.
\item Higher associativity relations hold for the composition of codimension $n$ junctions into junctions between defects $u_1(g_1), \cdots, u_1(g_1 g_2 g_3 \cdots g_{n+1})$.
The local equivalence of the composite junctions is implemented by $u_{n+1}(g_1,g_2,g_3,\cdots, g_{n+1})$.
\end{itemize}
In particular, all the defects, junctions, etc. are invertible.

Such a formidable collection of 
 associating defects is precisely what one would use to define abstractly a non-anomalous 
action of an internal symmetry $G$ on a local quantum field theory \cite{Gaiotto:2014kfa}. It can also be visualized as a map from 
the classifying space $BG$ to the space of continuous deformations of the gapped system. 

Conversely, suppose that we are given a system equipped with such a collection of defects. 
Then we claim that the system can be given an equivalent local lattice realization 
where the symmetry $G$ acts independently on the local degrees of freedom at each lattice site. 
We can build such a realization with the help of an auxiliary lattice sigma model with target $G$.
We will describe the construction for a statistical model; the construction for the quantum lattice system is analogous. 
We place the $G$ sigma model at the vertices of some mesoscopic lattice and ``fill in'' the cell complex dual to the mesoscopic lattice, by placing the 
original lattice model in the $(d+1)$-dimensional cells, the $u_1(g_1^{\mathstrut} g_2^{-1})$ defects at $d$-dimensional cells between lattice sites with labels 
$g_1$ and $g_2$, the $u_2(g_1^{\mathstrut} g_2^{-1}, g_2^{\mathstrut} g_3^{-1})$ defects at $(d-1)$-dimensional cells between lattice sites with labels 
$g_1$, $g_2$ and $g_3$, etc. 

We should also discuss the redundancies of this description. As we label domain walls, junctions, etc.\ by an equivalence class/phase 
$u_n(g_1, \cdots, g_n)$, we make an implicit choice of deformation path.
The difference between two choices can be encoded into invertible codimension $n+1$ defects $\epsilon_n(g_1, \cdots, g_n)$ on the $u_n(g_1, \cdots, g_n)$.
If we change our choice, we will have to re-label lower dimensional junctions. For example, 
the $u_2(g_1, g_2)$ interfaces will be modified to $\epsilon_1(g_1) \circ \epsilon_1(g_2) u_2(g_1, g_2) \circ \epsilon^{-1}_1(g_1 g_2)$.

We have discussed previously a basic observation about invertible theories: their defects are in one-to-one correspondence 
with lower dimensional theories, but the correspondence involves non-canonical choices of deformation paths. 
These choices affect the properties of junctions between these defects.  

If a fully canonical correspondence did exist, we could just identify the $u_n(\cdots)$ as invertible phases of the appropriate dimension 
and impose associativity of these phases under stacking. The classification of SPT phases would then be very simple. 
The group (under the stacking operation) of invertible phases in spacetime dimension $n$ is simply $\pi_0 \GP_n = \pi_{-n} \GP$.
The data $u_k(\dots)$ defines a group cochain valued in $\pi_{-n+k} \GP$, and if the correspondence played well with topological manipulations, then associativity would require $u_k$ to be a cocycle --- $\d u_k = 0$ --- and different $u_k$ would be equivalent if they differed by a coboundary.
In other words, if this (wrong) assumption held, then the set of SPT phases with symmetry $G$ would be simply the group cohomology $\prod_{k=1}^\infty \H^k(B G; \pi_{-n+k} \GP)$.
%
%

In order to understand the correct classification of SPT phases, it is useful to build the required structure a step at the time. 
First, we can look for a collection of $(d-1)+1$-dimensional invertible phases $u_1(g)$, which stack correctly. 
That means $u_1(g_1) \times u_1(g_2) \simeq u_1(g_1 g_2)$, i.e. \footnote{We are ignoring an important subtlety here. See Section \ref{sec:six}.} \ 
\begin{equation*}
\d u_1 = 0
\end{equation*} 

Next, we can pick some random collection of junctions between $u_1(g_1) \times u_1(g_2)$ and $u_1(g_1 g_2)$, for all 
group elements $g_1$ and $g_2$. We will then describe $u_2(g_1,g_2)$ as the $(d-2)+1$-dimensional invertible phases
which should be stacked on top of our reference choice in order to define the group action. Now, we can 
test associativity of junctions. Associativity for the reference junctions will fail. 

\begin{figure}[t]
\includegraphics[width=16cm]{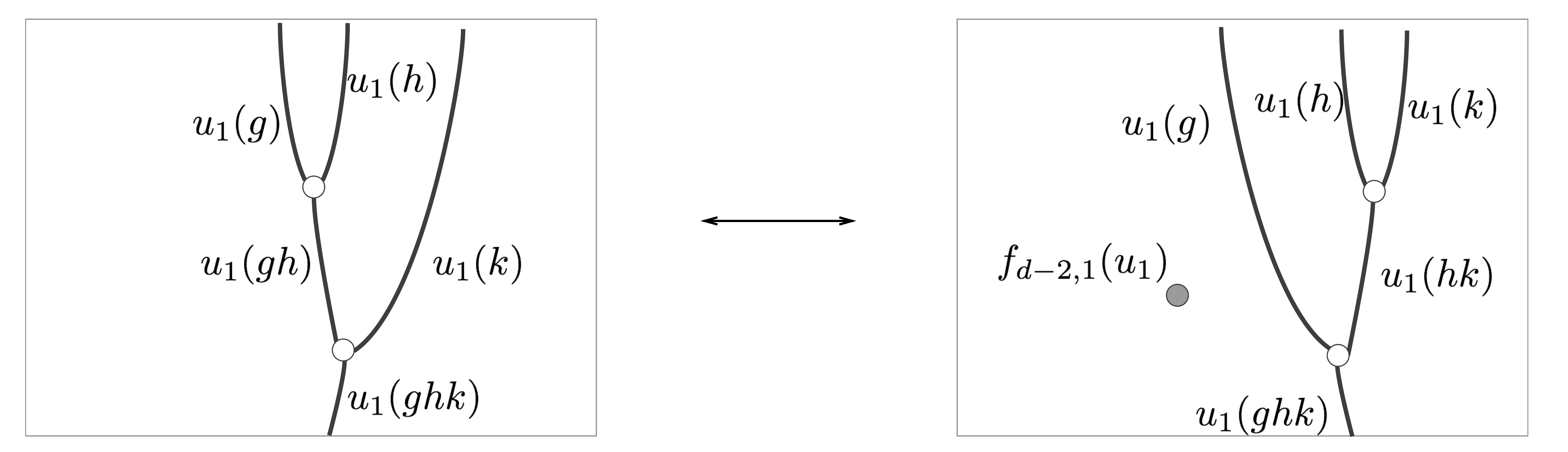}
\caption{Failure of associativity for reference junctions}\label{fig:eight}
\end{figure}

The failure is measured by the difference between two junctions between $u_1(g_1)$, $u_1(g_2)$, $u_1(g_3)$ and $u_1(g_1 g_2 g_3)$ interfaces and 
can be described as some $(d-2)+1$ dimensional phase $f_{d-2,1}(u_1)$. See Figure \ref{fig:eight}.
The $u_2(g_1,g_2)$ phases must be chosen in such a way to cure the lack of associativity of the reference phases. This means that the correct
condition takes the form
\begin{equation*}
\d u_2 = f_{d-2,1}(u_1)
\end{equation*} 

The same pattern persists for junctions of higher codimension. The constraints satisfied by the $u_k$ junctions 
receive corrections which depend on the $u_j$ with $j<k$, of the generic form
\begin{align*}
\d u_1 &= 0 \cr
\d u_2 &= f_{d-2,1}(u_1) \cr
\d u_3 &= f_{d-3,1}(u_2) + f_{d-3,2}(u_1) \cr
&\cdots \cr
\end{align*}

The redundancies of the description are deformed in a similar way:
\begin{align*}
u_2 &\to u_2 + \d \epsilon_1 \cr
u_3 &\to u_3 + \d \epsilon_2+ f_{d-3,1}(\epsilon_1) \cr
u_4 &\to u_4 + \d \epsilon_3+ f_{d-4,1}(\epsilon_2)+ f_{d-4,2}(\epsilon_1) \cr
&\cdots \cr
\end{align*}

 The functions $f_{p,i}$ are potentially-complicated functions $\C^k(B G; \GP_{p+i}) \to \C^{k+i+1}(B G; \GP_p)$.
Essentially by definition, this modified complex is the generalized cohomology theory associated to $\GP$, whose differentials keep track of the non-trivial identification between 
defects and theories in lower dimension, i.e.\ the homotopy equivalences $\Omega\GP_n \to \GP_{n-1}$. 

\subsection{Stable cohomology operations} \label{k-invariants}


By the main result of \cite{MR0214056}, the existence of a
 description of $\H^\bullet(BG;\GP)$ in terms of functions $f_{p,i}$ deforming the differential on $\prod_{k=1}^\infty \C^k(B G; \pi_{-n+k} \GP)$ is automatic: it holds with $\GP$ replaced by any spectrum whose homotopy groups vanish in degrees above some cut-off. That work also sheds some light on the type of functions $f_{p,i}$ that can appear. They are in general nonlinear. They are \define{natural transformations} in the sense that they do not depend on the choice of group $G$: if $r : X \to Y$ is any map of simplicial sets, the compositions $\C^k(Y; \GP_{p+i}) \overset{f_{p,i}}\to \C^{k+i+1}(Y; \GP_p) \overset r \to \C^{k+i+1}(X; \pi_0\GP_p)$ and $\C^k(Y; \GP_{p+i}) \overset r \to \C^k(X; \pi_0\GP_{p+i}) \overset{f_{p,i}}\to   \C^{k+i+1}(X; \pi_0\GP_p)$ agree. It follows in particular that if $u_k$ is a $k$-cochain, then
$f_{p,i}(u_k)(g_1,\dots,g_{p+i+1})$ is some function just of the values of $u$ on products of the $g_i$.

The whole differential has an ``upper triangular'' form 
$$ D\begin{pmatrix} u_n \\ \vdots \\ u_2 \\ u_1 \end{pmatrix}
= \begin{pmatrix}
 \d & -f_{0,1} & \dots & -f_{0,n-1} \\
  & \ddots & & \vdots \\
  && \d & -f_{n-2,1} \\
  &&& \d
\end{pmatrix}
\begin{pmatrix} u_n \\ \vdots \\ u_2 \\ u_1 \end{pmatrix}
$$
with the caveat that the functions $f_{p,i}$ are not linear. The requirement that $D^2 = 0$ implies a lot about the functions $f_{p,i}$. In particular, it implies that for each $p$, $f_{p,1}$ takes cocycles to cocycles and takes coboundaries to coboundaries. It gives therefore a \define{cohomology operation} $[f_{p,1}] : \H^k(-;\GP_{p+1}) \to \H^{k+2}(-;\GP_p)$. In fact, \cite{MR0214056} shows that on cohomology, each function $[f_{p,1}]$ \emph{is} linear, although it is represented nonlinearly at the cochain level. Furthermore, the functions $[f_{p,1}] : \H^k(-;\GP_{p+1}) \to \H^{k+2}(-;\GP_p)$ for different $k$ are closely related as follows: for any abelian groups $A$ and $B$, any cohomology operation $\H^N(-;A) \to \H^{N+j}(-;B)$ determines cohomology operations $\H^n(-;A) \to \H^{n+j}(-;B)$ for $n<N$, and a \define{stable} operation is a collection of operations, one for each $N \to \infty$, which each determine all the previous ones. The \define{degree} of a stable cohomology operation is just the degree by which it raises (it never lowers) cohomology classes; for example, $[f_{p,1}]$ is of degree~$2$. Degree-$i$ stable cohomology operations between abelian groups $A$ and $B$ are the same as degree-$i$ maps between the Eilenberg--MacLane spectra determined by $A$ and $B$.\footnote{The \define{Eilenberg--MacLane spectrum} for $A$, usually denoted $HA$, is the spectrum such that cohomology with coefficients in that spectrum is ordinary cohomology with coefficients in $A$.} The stable cohomology operations $[f_{p,1}]$ coming from the deformed differential on $\GP$ are called the \define{k-invariants} of the spectrum $\GP$.

The requirement $D^2 = 0$ implies further relationships between the $f_{p,i}$. For example, it implies that, if $u_k \in \Z^k(G; \pi_0\GP_{n-k})$ is an ordinary cocycle, then $\d f_{n-k-1,2}(u) = f_{n-k,1}(f_{n-k-1,1}(u))$ is a coboundary. In particular, the functions $f_{p,2}$ are typically not themselves cohomology operations --- they do not take cocycles to cocycles --- but
rather they are trivializations of compositions of $f_{p,1}$s. It follows that
 if the $f_{p,1}$s are given, then the data needed to define $f_{p,2}$ is a torsor over the space of stable cohomology operations. Similar results apply for $f_{p,i}$ with higher $i$. The functions $f_{p,2}$ are called the \define{j-invariants} of the spectrum $\GP$.

\subsection{SPT phases and reduced cohomology} \label{reduced cohomology}
In \S\ref{section symmetries and defects} we  used the definition of SPT phase as a phase which becomes trivial when the symmetry is neglected.
As a consequence we have associated them to  \emph{reduced} cohomology classes, built from 1-cochains, 2-cochains, etc., without any $0$-cochains. Given a spectrum $T$, we write $\H^\bullet(-;T)$ for the unreduced cohomology and $\tilde{\H}^\bullet(-;T)$ for reduced cohomology.

It is easy to extend the definition to allow for a non-trivial invertible phase in the absence of symmetry, by including a 0-cochain $u_0$. 
This would result in the unreduced cohomology valued in the same spectrum. 
If $G$ does not act on the spectrum $\GP$ itself (in particular, if $G$ includes no antiunitary symmetries), then
the $0$-cochain $u_0$ never participates in the remainder of the complex: one can stack the $G$-symmetric phase defined by an unreduced cocycle $(u_0,u_1,\dots,u_n)$ with the phase defined by $u_0^{-1} \in \C^0(G; \pi_n\GP) = \pi_0\GP_n$ equipped with the trivial $G$-action; the result of this stacking is an SPT phase in the reduced sense.

\subsection{Phases versus choices of realization}\label{u1 versus z}

The astute reader will have noticed that in the above discussion, we used the infinite product $\prod_{k=1}^\infty \H^k(B G; \pi_{-n+k} \GP)$ when it seems we only used cochains of dimensions $1,\dots,n$. The reason for this is a slight cheat in the exposition of \S\ref{section symmetries and defects}. The data of a $0$-spacetime-dimensional quantum system consists simply of its ``partition function'' --- a number --- so that the group of invertible $0$d quantum systems is $\bC^\times$. But, since $\bC^\times$ is connected, these systems are all in the same phase. Thus the space $\GP_0$ of $0$d gapped phases is not the group $\bC^\times$, but rather the corresponding topological space, homotopy equivalent to $S^1$: $\pi_0 \GP_0 = 0$ and $\pi_1 \GP_0 = \bZ$. In particular, the exposition of \S\ref{section symmetries and defects} assumed that the final ``associativity'' data for an $n$-dimensional SPT phase consisted of some $0$-dimensional defects associated to each $n$-tuple in $G$. But since $\pi_0 \GP_0 = 0$, this is no data at all. Rather, the final associativity data consists 
somewhat formally of some ``$-1$-dimensional defects'' --- classes in $\bZ = \pi_1\GP =$ ``$\pi_0 \GP_{-1}$'' --- associated to $(n+1)$-tuples in~$G$. 

When $G$ is a finite group\footnote{More general statements are available, for example when working with measurable cochains.} and $n>0$, there is no difference between $\H^n(BG; \bC^\times)$, $\H^n(BG;U(1))$, and $\H^{n+1}(BG;\bZ)$.
Thus the classification of SPT phases protected by a finite group is indifferent to the choice of whether to pretend that $\pi_0 \GP = \bC^\times$ or $U(1)$ and $\pi_{>0}\GP = 0$ or to use the correct spectrum, for which $\pi_0 \GP = 0$ and $\pi_1 \GP = \bZ$. However, the difference between $\bC^\times$-as-a-set and $\bC^\times$-as-a-topological-space does play an important role in understanding the ``$E_8$ phase''; c.f.\ \S\ref{section beyond cohomology}.

\subsection{Stacking of SPT phases}
A marvellous property of generalized cohomology theories is that they are linear, even though the 
extended differentials are polynomials at the level of cochains, 
and so do not preserve the usual group operation on cochains ---
 instead, one can always define
a (noncommutative, nonassociative) ``sum'' operation on cochains, which becomes commutative and associative at the level of cohomology~\cite{MR0214056}. 

Physically, the sum operation on cohomology realizes the stacking operation on SPT phases. 
We can recover the explicit cochain-level operation by the same physical considerations 
which we used to find the cohomology differential.

Concretely, the interfaces and defects which encode the product of two SPT phases are obtained by stacking the 
interfaces and defects which encode the two SPT phases. 
\begin{itemize}
\item At the top level, we simply stack the interfaces: if the parent SPT's have interfaces $u_1(g)$ and $u'_1(g)$, 
the new SPT will have interfaces $$u_1(g)+ u'_1(g).$$
\item Next, we need to stack junctions. The crucial subtlety is that stacking two reference junctions 
may not give a reference junction. Thus the new SPT will have junctions 
$$u_2(g_1,g_2)+ u'_2(g_1,g_2)+ s_{d-2}(u_1, u'_1).$$
\item Etcetera.
\end{itemize}
In this manner, we can compute all the maps which occur in stacking operations:
$$
(u+u')_n = u_n + u'_n + \sum_k s_{d-n}(u_*,u'_*)
$$

\section{Examples} \label{sec:five}

The previous section explained in principle how to analyze SPT phases in terms of cohomology valued in invertible phases. We now implement this procedure in various examples. We will talk about both bosonic and fermionic phases of matter. To indicate the difference, we will call the spectrum of bosonic invertible gapped phases $\bGP$ and the spectrum of fermionic invertible gapped phases $\fGP$.

\subsection{Bosonic SPT phases and standard group cohomology}
The standard construction of bosonic SPT phases by group cohomology can be immediately described in the language of topological defects.  For simplicity, let $G$ be a finite group, so that we can avoid the subtlety from \S\ref{u1 versus z}. 
Then the SPT phase described by $[\alpha] \in \H^{d+1}(BG;U(1))$ corresponds to a system of defects as in \S\ref{section symmetries and defects} which are completely trivial except at the  lowest possible dimension: 
the point junctions of codimension $d+1$ carry the phase $\alpha(g_1, \cdots, g_{d+1})$, a representative of the group cohomology class 
$[\alpha]\in \H^n(B G; U(1))$.

Notice that ``point'' here means a point in space and time. For example, for an SPT phase in $1+1d$, these would be point-like junctions 
between domain walls for group elements $g$, $h$, $gh$. In order to ``measure'' $\alpha(g_1, \cdots, g_{d+1})$, one has 
execute the following operations:
\begin{enumerate}
\item Draw the configuration of domain walls in space-time around a single junction point and 
take a space slice before and after the point. 
\begin{itemize}
\item In $1+1d$ the two configurations would consist of a chain decorated either with a $g_1$ and an $g_2$ domain walls, well separated, or by a $g_1 g_2$ domain wall. 
\item In $2+1d$ the two configurations would consist of the two different ways of connecting a $g_1$, a $g_2$ and a $g_3$ domain walls 
to a $g_1 g_2 g_3$ domain wall, either by fusing the first two walls into a $g_1 g_2$ wall and then fusing it with the third wall, or fusing the last two walls into a $g_2 g_3$ wall and then fusing it with the first. 
\item Etcetera
\end{itemize}
\item Pick a specific identification between the ground-state of the theory decorated by domain walls 
in space and the trivial wavefunction. The identification must be local: the details of the identification near one domain wall 
cannot depend on the existence of another domain wall far from it.  
\item The phases in $[\alpha]$ are simply the composition of the maps from the trivial wavefunction 
to the ground-states of the two configurations of domain walls and the gauge transformation relating the two 
configurations.%
\footnote{The basic reason the $[\alpha]$  can be non-trivial is that the gauge transformation relating the two configurations 
of domain walls is less ``local'' than the identifications with the trivial phase: the gauge transformation 
at a point depends on the whole configuration of domain walls.}
\end{enumerate}
Restoring from $U(1)$ to $\bZ$, we find that the spectrum $\bGP$ of bosonic invertible gapped phases has $\pi_1 \bGP = \bZ$, $\pi_0 \bGP = 0$, and, since there are no nontrivial $0+1$d bosonic gapped phases, $\pi_{-1} \bGP = 0$.

\subsection{Bosonic SPT phases beyond group cohomology}\label{section beyond cohomology}

The classification of bosonic SPT phases by the standard group cohomology will work only up to the dimension where non-trivial invertible bosonic phases 
with no symmetry first appear. The nature of these phases depends rather strongly on the precise setup. Physically, there could be differences between 
invertible phases of matter in the condensed matter sense and invertible phases of gapped relativistic quantum field theories. 
Mathematically, definitions based on traditional or extended TFTs, framed, unframed or partially framed TFTs or 
cobordism groups will give different answers, which in turn can be different from the classification of 
invertible phases of matter in the condensed matter sense or may coincide with it for no obvious reason. 

The first non-trivial bosonic invertible phase in the condensed matter  sense appears in $2+1d$: it is Kitaev's $E_8$ phase. 
This phase is characterized by the emergence at the boundary of edge modes of chiral central charge $c=8$, in the form of an $E_8$ WZW model at level $1$. 
It is generally expected that any $n$-th power of this phase should be non-trivial and support edge modes of central charge $8n$. For evidence that the $E_8$ phase has infinite order, see \S\ref{time reversal E8}.

Thus the spectrum $\bGP$ should have homotopy groups 
$$ \pi_1 \bGP = \bZ, \quad \pi_0 \bGP = 0, \quad \pi_{-1} \bGP = 0, \quad \pi_{-2} \bGP = 0, \quad \pi_{-3} \bGP = \bZ, \quad \dots $$
(As discussed in \S\ref{u1 versus z}, the $\bC^\times$ or $U(1)$ worth of $0$d quantum systems becomes the space $\bGP_0 = S^1$ with $\pi_0 \bGP_0 = 0$ and $\pi_1 \bGP_0 = \bZ$.)
These homotopy groups match the homotopy groups of the spectrum $\Sigma I_\bZ MSO$, the Anderson dual to oriented bordism.%
\footnote{The \define{oriented bordism spectrum} $MSO$, also called $\Omega^{SO}$, is the connective spectrum whose $k$-cells are $k$-dimensional oriented cobordisms; in particular, $\pi_k MSO$ is the group of oriented $k$-manifolds up to cobordism. 
\define{Anderson duality} is a type of duality for spectra related to Pontryagin duality for abelian groups. Let us mention one property of it.
Suppose that $T$ is a spectrum whose homotopy groups $\pi_k T$ are finitely generated as abelian groups. Then for each $k$, we can noncanonically split $\pi_k T \cong (\pi_k T)[\mathrm{free}] \oplus (\pi_k T)[\mathrm{tor}]$, where $(\pi_k T)[\mathrm{free}] \cong \bZ^r$ is a free abelian group and $(\pi_k T)[\mathrm{tor}]$ is the torsion subgroup of $\pi_k T$. The shifted Anderson dual $\Sigma I_\bZ T$ of $T$ then has the following homotopy groups:
$$ \pi_{-k} \Sigma I_\bZ T \cong \hom\left((\pi_{k+1} T)[\mathrm{free}],\bZ\right) \oplus \hom\left((\pi_k T)[\mathrm{tor}],U(1)\right) \cong (\pi_{k+1} T)[\mathrm{free}] \oplus (\pi_k T)[\mathrm{tor}]. \quad \text{(noncanonical!)}$$
The more canonical statement describes $\pi_{-\bullet}\Sigma I_\bZ T$ in terms of Ext groups $\operatorname{Ext}^\bullet(\pi_\bullet T, \bZ)$.
}
 We believe that in fact $\bGP = \Sigma I_\bZ MSO$, or at the very least the spectrum $\bGP_{\leq 3}$ of bosonic phases of dimension at most $3$ agrees with the $(-3)$-connective cover $\Sigma I_\bZ MSO\langle -3\rangle$. This belief can be checked by calculating $k$-invariants on both sides. Details will appear elsewhere.

One may wonder if any other invertible phases may occur in dimension higher than $2+1d$. In a TFT setup, 
one has $4d$ Crane--Yetter models built from Modular Tensor Categories, but they are expected to be trivial 
as (Walker-Wang) condensed matter phases, for the same reason that allows MTC's to appear as 
categories of anyons in $2+1d$ lattice systems.
In fact, the partition function on a $4$-manifold $M$ for a Crane--Yetter model described by a MTC with central charge $c$ is $\exp(\operatorname{signature}(M)\,2\pi i c /8)$, and it is reasonable to believe that the full space of $4d$ invertible TFTs consists of this $U(1)$ worth of partition functions. Notice that for finite $G$, $\H^n(G; \bZ) = \H^{n-1}(G; U(1))$ when $n>1$. Thus a spectrum with $U(1)$ worth of phases in dimension $3+1d$ would give the same classification of SPT phases in  sufficiently high dimension 
as a spectrum with a $\bZ$ worth of phases in dimension $2+1d$; compare \S\ref{u1 versus z}. From this perspective, the $2+1d$ $E_8$ phase is nothing but the \emph{path} that wraps one around the $U(1)$ worth of $3+1d$ TFTs;
compare the paths used in \S\ref{definition of spectra} to identify $\Omega \GP_n$ with $\GP_{n-1}$.

\subsection{Restricted supercohomology}\label{supercohomology section}

We turn now to fermionic phases of matter. There is quite a lot of debate about what the spectrum of fermionic phases is:
\cite{Kapustin:2014dxa}  conjectures that the spectrum of fermionic phases is $\Sigma I_\bZ MSpin$,%
\footnote{Parts of the paper \cite{Kapustin:2014dxa} use just  the torsion part of the homotopy groups of $\Sigma I_\bZ MSpin$; as observed in \cite{Kapustin:2014dxa,FH}, the non-torsion part seems to describe phases of matter that do not correspond to truly-topological field theories. 
Since in this paper we do not assume any \emph{a priori} connection to topological field theory, we will not see such a distinction, and we will see non-torsion groups of $\Sigma I_\bZ MSpin$  in \S\ref{beyond majorana}.}
whereas in a field theory context \cite{FH} uses $\Sigma I_\bZ S$, where $S$ denotes the sphere spectrum.%
\footnote{The paper \cite{FH}  supersedes the earlier paper \cite{DFreed}, which suggests more strongly that $\Sigma I_\bZ S$ is the spectrum of fermionic phases. The main result of \cite{FH} is that ``reflection positive'' invertible phases of TFTs are classified by a spectrum of shape $\Sigma I_\bZ MH$, where $H$ is some to-be-determined ``tangential structure,'' for example $H = SO$ or $H = Spin$.
In particular, \cite{FH} would support the prediction $\Sigma I_\bZ MSpin$ if fermionic=Spin.}

However, the low-dimensional homotopy groups are undebatable. Letting $\fGP$ denote the spectrum of fermionic phases, we have
$$ \pi_1 \fGP =\bZ, \quad \pi_0 \fGP = 0, \quad \pi_{-1} \fGP  = \bZ_2, \quad \pi_{-2} \fGP  = \bZ_2. $$
The homotopy groups in degrees $1$ and $0$ are as in \S\ref{u1 versus z}. The nontrivial $0+1d$ phase  is the fermion, and the nontrivial  $1+1d$ phase  is the Majorana chain. These homotopy groups are consistent with both proposals $\fGP = \Sigma I_\bZ S$ and $\fGP = \Sigma I_\bZ MSpin$.

Let us consider first the connective cover $\fGP\langle -1\rangle = \fGP_{\leq 1}$, whose only nontrivial homotopy groups are $\pi_1 = \bZ$ and $\pi_{-1} = \bZ_2$. A spectrum with only two nontrivial homotopy groups is completely determined by (those homotopy groups and) the k-invariant. In this case, there are exactly two possibilities: the trivial k-invariant, and a unique nontrivial one. Let $\Box : \H^\bullet(-;\bZ_2) \to \H^{\bullet+1}(-;\bZ)$ denote the \define{Bockstein homomorphism} for the extension $0 \to \bZ \overset{\times 2}\to \bZ \to \bZ_2 \to 0$. (If $G$ is a finite group, then $\H^{\bullet+1}(BG;\bZ)$ and $\H^\bullet(BG; \bR/\bZ)$ are canonically identified, and this Bockstein homomorphism is simply the inclusion $\frac12 : \bZ_2 \to \bR/\bZ$.) The unique nontrivial degree-3 stable cohomology operation from $\bZ_2$ to $\bZ$ is the composition $$\H^\bullet(-;\bZ_2) \overset{\Sq^2}\longto \H^{\bullet+2}(-;\bZ_2) \overset\Box\longto \H^{\bullet+3}(-;\bZ),$$
where $\Sq^2 $ is the second Steenrod square.

In particular, to compute $\fGP_{\leq 1}$, it suffices to show that it is not the trivial extension, as then it must be the nontrivial extension with k-invariant $\Box \circ \Sq^2$. One may show this in many ways. The simplest is to recognize that the generalized cohomology theory
associated to $\fGP_{\leq 1}$ describes systems where 0-dimensional junctions of domain walls carry phases and 1-dimensional junctions carry vector spaces of even or odd fermion number. These are precisely the systems considered in the seminal paper \cite{Gu:2012ib}. 

Thus cohomology with coefficients in $\fGP_{\leq 1}$ is nothing but the \define{(restricted) supercohomology} of \cite{Gu:2012ib,KitaevTalk2}. The construction and classification of spectra in terms of stable cohomology operations from \S\ref{k-invariants} 
explains the appearance of $\Sq^2$ in the formulas for supercohomology.
In summary, we can think of the spectrum $\fGP_{\leq 1}$ as a ``complex'' with two terms, $\bZ$ and $\bZ_2$, and differential
$$ D = \begin{pmatrix} \d & \Box \Sq^2 \\ 0 & \d \end{pmatrix}.$$


\subsection{Extended supercohomology}\label{extended supercohomology}

We now consider the spectrum $\fGP_{\leq 2}$ with homotopy groups $\bZ, 0, \bZ_2,\bZ_2$ in degrees $1$, $0$, $-1$, and $-2$. We have computed already that the k-invariant connecting the $\bZ_2$ worth of $0+1d$ phases with the $\bZ$ in top degree is $\Box\circ \Sq^2$. To compute the spectrum, we must compute the k-invariant connecting the two $\bZ_2$s and also the j-invariant trivializing the composition of the two k-invariants. The k-invariant connecting the two $\bZ_2$s is a degree-$2$ stable cohomology operation from $\bZ_2$ to $\bZ_2$, of which there is a unique nontrivial one: $\Sq^2$.

Let us temporarily assume that we have shown that the k-invariant connecting $\pi_{-2}\fGP = \bZ_2$ and $\pi_{-1} \fGP = \bZ_2$ is nonzero, hence $\Sq^2$. We will argue that this is enough to fully determine the spectrum $\fGP_{\leq 2}$. More specifically, we claim that up to noncanonical equivalence, there is a unique spectrum whose only nonzero homotopy groups are $\bZ, 0, \bZ_2,\bZ_2$ in degrees $1$, $0$, $-1$, and $-2$ and such that both k-invariants are nonzero.  

Indeed, the remaining data of such a spectrum should be a trivialization of the composition of the two k-invariants. One can see that such a trivialization exists by using the Adem relations $\Sq^2\Sq^2 \simeq \Sq^3\Sq^1$ and $\Sq^3 \simeq \Sq^1\Sq^2$ and by noting that $\Box  \Sq^1 \simeq 0$, and so $\Box \Sq^2\Sq^2$ is trivial in cohomology. There is still the choice of particular j-invariant: the set of inequivalent choices is a torsor for the group of stable cohomology operations of the correct degree --- in our case, degree $2+3-1 = 4$ --- from $\pi_{-2}\fGP = \bZ_2$ to $\pi_1\fGP = \bZ$. There is precisely one non-zero such operation, namely $\Box \Sq^2\Sq^1$, and so there are precisely two possible j-invariants, differing by that operation. With a choice of j-invariant made, the spectrum $\fGP_{\leq 2}$ is determined.

So our claim amounts to the claim that the two j-invariants lead to equivalent spectra, which will in turn boil down to the fact that $\Box \Sq^2 \Sq^1$ factors through the k-invariant $\Box \Sq^2$ connecting $\pi_{-1}\fGP = \bZ_2$ with $\pi_1 \fGP = \bZ$.  Indeed, let $j$ denote one of the two $j$-invariants, determining the spectrum with differential
$$ \begin{pmatrix} \partial & \Box\Sq^2 & j \\ & \partial & \Sq^2 \\ && \partial \end{pmatrix}.$$
The other possible spectrum has differential like above but with $j$ replaced by $j + \Box \Sq^2 \Sq^1$. But:
$$ \begin{pmatrix} \partial & \Box\Sq^2 & j + \Box\Sq^2\Sq^1 \\ & \partial & \Sq^2 \\ && \partial \end{pmatrix} = \begin{pmatrix} \id && \\ &\id & \Sq^1 \\ && \id \end{pmatrix} \begin{pmatrix} \partial & \Box\Sq^2 & j \\ & \partial & \Sq^2 \\ && \partial \end{pmatrix} \begin{pmatrix} \id && \\ &\id & \Sq^1 \\ && \id \end{pmatrix}^{-1} $$
The matrix $\left(\begin{smallmatrix} \id && \\ &\id & \Sq^1 \\ && \id \end{smallmatrix}\right)$ thus describes a spectrum equivalence between the two possible j-invariants.\footnote{One can compare this to the case of breaking up $\bZ_8$ as an extension $\bZ_2\cdot\bZ_2\cdot\bZ_2$. The k-invariants correspond to deciding that both $\bZ_2\cdot\bZ_2$s should compile to $\bZ_4$s, i.e.\ that there should be a ``carry'' so that $01 + 01 = 10$ and not $00$.  This gives the following rules in the putative $\bZ_8$: $100 + 100 = 000$, $010 + 010 = 100$, and $001 + 001 = {x}10$. The j-invariant corresponds to the choice of whether $x = 0$ or $1$. The two choices give isomorphic groups, where the isomorphism interchanges $001$ and $011$. The isomorphism above involving $\Sq^1$ is analogous.}

So to complete the calculation of $\fGP_{\leq 2}$, it suffices to show that both k-invariants are nonzero. There are many ways to do this. For example, one can observe that there is a $2+1d$ fermionic SPT phase protected by $\bZ_2$ of order~$8$ under stacking \cite{GuLevin,SPT9}. But direct computation of $\H^3(\bZ_2; T)$ for the different possible spectra $T$ with the same homotopy groups as $\fGP_{\leq 2}$ shows that the only way to get an order-$8$ element is when both k-invariants are nonzero.

Cohomology with coefficients in $\fGP_{\leq 2}$ is precisely the \define{extended supercohomology} of \cite{SPT8,SPT11}.

\subsection{Algebraic description of extended supercohomology}\label{super algebra}

There is another way to compute the spectra $\fGP_{\leq 1}$ and $\fGP_{\leq 2}$: we can realize them algebraically. The category of all $0+1d$ fermionic gapped phases of matter, not necessarily invertible, is the category $\SV_\bC$ of complex supervector spaces, and the spectrum $\fGP_{\leq 1}$ of invertible fermionic phases is simply (the spectrum built from) the category $\SV_\bC^\times$ of super lines. This is clear from the description: the nontrivial $0+1d$ invertible phase \emph{is} the odd line --- the fermion. The statement that the k-invariant is $\Box\Sq^2$ and not zero is simply the statement of the Koszul sign rule: the fermion braids with itself for a sign rather than trivially.

Similarly, $\fGP_{\leq 2}$ is nothing but (the spectrum built from) the symmetric monoidal bicategory $\SA_\bC^\times$ of Morita-invertible complex superalgebras, super Morita equivalences, and intertwiners. The nontrivial object in $\SA_\bC^\times$ is, from this perspective, the Clifford algebra $\Cliff(1)$. The  claim about the k-invariant connecting the two $\bZ_2$s, then, is simply the claim that the self-braiding of the object $\Cliff(1) \in \SA_\bC$ is isomorphic to the parity-reversal of the identity bimodule on $\Cliff(1) \otimes \Cliff(1)$, and not to the identity bimodule. We check this explicitly:

 Let the generators of $A = \Cliff(1) \otimes \Cliff(1)$ be $x$ and $y$, subject to the relations $x^2 = y^2 = -1$ and $xy = -yx$. The identity bimodule on $A$ is $A$ itself treated as a bimodule. To distinguish, we will call the bimodule $M$, and its basis $1_M$, $x_M$, $y_M$, and $(xy)_M$. Of course, the basis vectors labeled $1$ and $(xy)$ are even, and those labeled $x$ and $y$ are odd. It is acted on from both the left and the right by $A$ with the obvious actions. For example, $x \triangleright x_M = -1_M$, $y_M \triangleleft x_R = -(xy)_M$, etc.

The self-braiding of $\Cliff(1)$ is another bimodule between $A$ and itself, which we will call $N$. Its underlying supervector space is the same as that of $M$ --- we will indicate the basis of $N$ by $1_N$, $x_N$, $y_N$, and $(xy)_N$. The right action of $A$ on $N$ is the same as that of $A$ on $M$. But the left action of $A$ is ``braided'' in the sense that left actions of $x$ and $y$ are reversed. For example, $x \triangleright x_N = -(xy)_N$, since $y \triangleright x_M = -(xy)_M$.

One may now check that the odd map $1_M \mapsto (x_N + y_N)/\sqrt{2}$ is an odd unitary isomorphism between the bimodules $M$ and $N$. 
Indeed, $1_M$ and $(x_N + y_N)/\sqrt{2}$ are the unique-up-to-phase elements $v$ in their respective modules with the property that $a \triangleright v = v \triangleleft a$ for all $a \in A$.
 This completes the proof that the k-invariant connecting $\pi_{-2}\SA_\bC^\times = \bZ_2$ and $\pi_{-1}\SA_\bC^\times = \bZ_2$ is $\Sq^2$.

\subsection{Beyond the Majorana layer}\label{beyond majorana}
In dimension $2+1d$, there is expected to be an invertible fermionic phase, a massive Majorana fermion, 
whose edge mode is a chiral fermion, with central charge $c=\frac12$. This plays a role analogous to the 
$E_8$ phase in \S\ref{section beyond cohomology}. 

We thus expect $$ \pi_1 \fGP =\bZ, \quad \pi_0 \fGP = 0, \quad \pi_{-1} \fGP = \bZ_2, \quad \pi_{-2} \fGP = \bZ_2, \quad \pi_{-3} \fGP = \bZ.$$
These groups are compatible with the prediction that $\fGP = \Sigma I_\bZ MSpin$ but incompatible with the prediction that $\fGP = \Sigma I_\bZ S$.

We will not compute here the k- and j-invariants connecting the $\bZ = \pi_{-3}\fGP$ to the remaining homotopy groups. We do observe that the k-invariant connecting that $\bZ$ to the $\bZ_2 =  \pi_{-2} \fGP$ must be nontrivial: if it were trivial, then all elements of the group $\H^4(\bZ_2^T;\fGP)$ classifying time-reversal fermionic $3+1d$ phases would have order $\leq 8$, but it is known that in fact there is a $\bZ_{16}$ classification \cite{Kit16,FCV}.

\subsection{A spectrum of SPT phases}
It is worth pointing out that SPT phases for some symmetry $H_b$ are themselves invertible phases of matter. 
This follows immediately from the fact that generalized cohomology groups are groups. Conversely, an invertible phase of matter 
with symmetry  $H_b$ is also an $H_b$ SPT phase. 

The space of invertible systems with symmetry $H_b$ defines a spectrum, which we can denote as $\GP(H_b)$. 
This is just the spectrum of maps from $BH_b$ to $\GP$. 

If we take an SPT phase with symmetry $G_b \times H_b$ and couple it to a $G_b$ flat connection, 
we will be left with a collection of domain walls for $G_b$ valued in $\GP(H_b)$. This collection has the same information as 
the usual collection of $G_b \times H_b$ domain walls valued in $\GP$. Indeed, almost by definition one has 
$$
\H^\bullet(BG_b \times BH_b;\GP) = \H^\bullet(BG_b;\GP(H_b)).
$$

Note, however, that by definition an \define{SPT phase} with symmetry $G_b$ is not just a $G$-symmetric invertible phase, encoded by a map $BG_b \to \GP$, but one that trivializes when the $G_b$ symmetry is forgotten; as in \S\ref{reduced cohomology}, the set of SPT phases is the \emph{reduced} cohomology  $\tilde \H^\bullet(BG_b; \GP)$, and that in general
$$ \tilde\H^\bullet(BG_b \times BH_b; \GP) \neq \tilde\H^\bullet(BG_b; \GP(H_b)).$$
Indeed, with the definition of SPT phase in terms of {trivial} phases with $G_b$-symmetry, the spaces of SPT phases of varying dimensions do not satisfy the relation between defects and deformations from \S\ref{defects} and so do not form an $\Omega$-spectrum as in \S\ref{definition of spectra}.

\section{Categorical actions and antiunitarity}\label{sec:six}

Our analysis so far applies to situations where the domain walls for the symmetry group elements in an SPT phase
can be fully identified with lower dimensional phases. In particular, the domain wall network 
is recast as a network of lower-dimensional systems, within which one does all subsequent calculations. 

This assumption needs to be relaxed if we want to deal with situations such as symmetry groups which act 
in an anti-unitary (aka time-reversal) manner. 
It is possible to define a domain wall for an anti-unitary symmetry, as described in footnote \ref{foot}:
the parts of the system on the two sides of such a domain wall are assembled together 
by a twisted tensor product. 

In an SPT phase, such a domain wall may support a lower dimensional invertible phase, 
but it will have an additional property: it  transforms any lower-dimensional physical system carried across it. 
For example, a local operator of multiplication by a complex number will get 
conjugated if transported across the wall. An invertible $2+1d$ $E_8$ phase 
will be conjugated to the opposite phase $\bar E_8$, etc.  

In general, we should allow for a situation where the domain walls for the group symmetry 
can be described as a lower dimensional phase stacked on top of a reference domain wall which acts non-trivially 
on the space of physical systems we consider and, in particular, on the space of invertible systems. 
This action will modify the associativity relations produced by our analysis of SPT phases. 

Mathematically, this action translates into a {\it categorical action} of the symmetry group $G_b$ onto 
the spectrum of invertible phases $\GP$. The categorical action collects all the data of how the 
reference domain walls act on invertible phases, including all the lower-dimensional domain walls and junctions 
associated to the intersection between invertible phases and reference domain walls. 

Such a categorical action allows one to define a \define{twisted} version of generalized cohomology,
which will control the classification of SPT phases.

Geometrically, the categorical action describes a fibration over $BG_b$ with fiber $\GP$.
 The twisted generalized cohomology groups are the homotopy 
groups of the space (spectrum) of sections of such a fibration, rather than the space of maps from $BG_b$ to $\GP$ \cite{KitaevTalk}. 

Another situation where categorical group actions apply involves fermionic SPT phases where the overall symmetry group $G_f$ 
is not the direct product of $G_b$ and fermion parity. We can still label the invertible theories which arise at 
domain walls by $G_b$ group elements, but they fuse up to extra $(-1)^F$ fermion parity walls. The mismatch is
controlled by the $\bZ_2$-valued $G_b$ cocycle $n_2(g_1,g_2)$ defining the extension $G_f$. 

The $(-1)^F$ walls are invisible to 0-dimensional bosonic local operators, but act non-trivially on the higher-dimensional 
invertible phases. For example, the intersection of $(-1)^F$ walls with the 1-dimensional 
lines which carry fermion number $1$, i.e.\ the non-trivial $0+1d$ invertible fermionic phase,
produces an extra factor of $-1$. Similar considerations apply for the intersection 
of $(-1)^F$ walls with other invertible fermionic phases. This can be translated into a categorical action relating 
different parts of $\fGP$. 

The remainder of this section studies basic examples of anti-unitary SPT phases in terms of twisted cohomology.

\subsection{Time-reversal and ordinary group cohomology} \label{twisted ordinary cohomology}
If we focus on bosonic theories and ignore the existence of non-trivial invertible phases, we can 
describe in our language the standard classification of SPT phases for anti-unitary symmetries 
as twisted group cohomology. The categorical action of time-reversal symmetry reduces to the 
action of complex conjugation on phases. 

Passing from phases to integers in one degree lower, this is an action of $G$ on $\bZ$ itself. The group of automorphisms of $\bZ$ is isomorphic to $\bZ_2$: the action of $x \in \bZ_2$ is of course by multiplication by $(-1)^x$. Suppose that $G$ is equipped with a homomorphism $\epsilon : G\to \bZ_2$, providing an action of $G$ on $\bZ$ by multiplication by $(-1)^{\epsilon(g)}$. Concretely, 
$g \in G$ acts antiunitarily whenever $\epsilon(g) \neq 0$. 

The effect of the categorical action is to replace the usual cohomology $\H^{n+1}(BG;\bZ)$ by the \define{twisted} cohomology $\H^{n+1}(BG;\bZ^\epsilon)$. Recall that the cochains for twisted cohomology are the same as those for untwisted cohomology, but the twisted differential is
\begin{multline*}
\d_\epsilon(\alpha)(g_0,\dots,g_n) = (-1)^{\epsilon(g_0)}\alpha(g_1,g_2,\dots,g_n) - \alpha(g_0g_1,g_2,\dots,g_n) + \alpha(g_0,g_1g_2,\dots,g_n)\\
- \dots \pm \alpha(g_0,\dots,g_{n-2},g_{n-1}g_n) \mp \alpha(g_0,\dots,g_{n-2},g_{n-1}),\end{multline*}
differing from the untwisted differential only in the first term.

In particular, consider the case when $G = \bZ_2$ and $\epsilon : G \to \bZ_2$ is the identity. Then the twisted group cohomology is
$$ \H^{n+1}(\bZ_2; \bZ^\epsilon) = \begin{cases} \bZ_2, & n \text{ even (and positive),} \\ 0, & n \text{ odd,} \end{cases}$$
giving a proposed spectrum $\bGP(\bZ_2^T)$ of ``time-reversal protected phases'' with homotopy groups $\pi_{0,-1,-2,\dots} = \bZ_2, 0, \bZ_2, 0 ,\dots$.

The $\bZ_2 = \pi_0 \bGP(\bZ_2^T)$ consists of the two possible signs of a real $0$-dimensional partition function. An interesting case is in spacetime dimension $2$. Then the $\bZ_2 = \pi_{-2}\bGP(\bZ_2^T)$ worth of $1+1d$ time-reversal-protected invertible systems consists of the trivial system and the Haldane chain.

$\H^{n+1}(BG;\bZ^\epsilon)$ provides the complete classification of antiunitary SPT phases in spacetime dimensions $n\leq 2$. But in high dimensions  it is incomplete, since according to \S\ref{section beyond cohomology} the bosonic spectrum $\bGP$ is not just the Eilenberg--MacLane spectrum of ordinary cohomology. Said another way, the correct spectrum $\bGP(\bZ_2^T)$ of bosonic time-reversal-protected phases does have homotopy groups
$$ \pi_0 \bGP(\bZ_2^T) = \bZ_2 ,\quad \pi_{-1} \bGP(\bZ_2^T) = 0 , \quad \pi_{-2} \bGP(\bZ_2^T) = \bZ_2,$$
but after this the homotopy groups become more complicated. 

Comparing with \S\ref{super algebra}, we can identify the bosonic spectrum $\bGP(\bZ_2^T)_{\leq 2}$ with the spectrum of Morita-invertible real (bosonic) algebras:
the Haldane chain corresponds to the quaternion algebra~$\bH$.
As in \S\ref{u1 versus z}, when $G$ is finite we may identify $\H^{n+1}(-;\bZ^\epsilon)$ with $\H^n(-;U(1)^\epsilon)$.

\subsection{Time reversal and $E_8$}\label{time reversal E8}

Including the $E_8$ phase as in \S\ref{section beyond cohomology}  leads to  a classification of SPT phases in terms of pairs of cochains of degrees differing by $4$. The first non-trivial example arises in $3+1d$, for time-reversal symmetry.
This  happens because $\H^1(\bZ_2;\bZ^\epsilon) = \bZ_2$: the domain walls which implement time-reversal symmetry 
may support an $E_8$ phase. Examples of proposed classifications of $3+1d$ bosonic SPT phases suggest that 
the $k$-invariant should restrict trivially here, so that the generalized cohomology theory collapses to 
$\H^1(G;\bZ^\epsilon) \oplus \H^5(G;\bZ^\epsilon)$.

Notice that this answer depends critically on powers of the $E_8$ phase being non-trivial. For example, the triple power 
$E_8^3$ is undetectable by traditional TFT means as it has partition function~$1$ on all manifolds. Nevertheless, 
$\H^1(\bZ_2;\bZ^\epsilon)$ is very different from $\H^1(\bZ_2;\bZ_3^\epsilon)$!

For unitary symmetries, the first non-trivial possibility of beyond-cohomology SPTs involving the $E_8$ phase occurs in $4+1d$, 
as $\H^2(G; \bZ)$ can be non-trivial. This would be a situation where $E_8$ phases appear at junctions of $G$-symmetry domain walls.

\subsection{Restricted Gu--Wen phases with time reversal} \label{gu wen time}

We turn now to the question of antiunitary actions in the presence of fermions. We discuss first the restricted supercohomology from~\S\ref{supercohomology section} corresponding to the spectrum $\fGP_{\leq 1}$ with homotopy groups
$$ \pi_1 \fGP_{\leq 1} = \bZ, \quad \pi_0 \fGP_{\leq 1} = 0, \quad \pi_{-1} \fGP_{\leq 1} = \bZ_2$$
and k-invariant $\Box \Sq^2$. As discussed in \S\ref{super algebra}, we may identify $\fGP_{\leq 1}$ with the spectrum $\SV_\bC^\times$ of invertible objects (and invertible morphisms) in the category of complex supervector spaces.

In the presence of fermions, there are two inequivalent notions of ``time reversal'': one can declare that $T^2 = 1$ or that $T^2 = (-1)^F$.
 These correspond to two different ways of extending ``complex conjugation'' to the category $\SV_\bC$. These two actions restrict to the two ways that $\bZ_2$ can act on $\fGP_{\leq 1}$ for which the induced action on $\pi_1\fGP_{\leq 1} = \bZ$ is nontrivial. 
Recall from \S\ref{supercohomology section} that $\fGP_{\leq 1}$ looks approximately like $\bZ \oplus \bZ_2$ but with differential $\bigl( \begin{smallmatrix} \d & \Box \Sq^2 \\ 0 & \d \end{smallmatrix} \bigr)$. In terms of this description, the two $\bZ_2$-actions on $\fGP_{\leq 1}$ are
$$ \begin{pmatrix} -\id & 0 \\ 0 & \id \end{pmatrix} \quad \text{or} \quad \begin{pmatrix} -\id & \Box \Sq^1 \\ 0 & \id \end{pmatrix}.$$

Suppose that $G$ is a finite group acting on $\fGP_{\leq 1}$ via $\epsilon : G \to \bZ_2$. Let $\d_\epsilon$ denote the twisted differential for ordinary cohomology discussed in \S\ref{twisted ordinary cohomology}. Regardless of the twisting, a degree-$n$ cochain for twisted $\fGP_{\leq 1}$ cohomology consists of an $(n+1)$-cochain $\alpha$ valued in $\bZ$ and an $(n-1)$-cochain $\beta$ valued in $\bZ_2$. The first action of $\bZ_2$ on $\fGP_{\leq 1}$ corresponds to the twisted differential
$$ \d_\epsilon(\alpha,\beta) = \bigl(\d_\epsilon(\alpha) + \Box \Sq^2 \beta, \d \beta\bigr).$$
The second one gives instead the twisted differential
$$ \d'_\epsilon(\alpha,\beta) = \bigl( \d_\epsilon(\alpha) + \Box \Sq^2 \beta + \Box(\epsilon \cup \Sq^1\beta), \d\beta\bigr).$$

Let us take $G = \bZ_2$ and $\epsilon = \mathrm{id}$ and compute the two possible spectra $\fGP(\bZ_2^T)_{\leq 1}$ of fermionic time-reversal-protected phases.
The homotopy groups of $\fGP(\bZ_2^T)_{\leq 1}$ are precisely the twisted cohomology groups for $G = \bZ_2$ and differentials $\d_\epsilon$ or $\d'_{\epsilon}$. In both cases we find $\pi_{\geq 1} \fGP(\bZ_2^T) = 0$ and
$ \pi_0 \fGP(\bZ_2^T) = \bZ_2.$
But the two differentials give different classifications of invertible $0+1d$ systems.
For the differential called $\d_\epsilon$ above, without the $\Sq^1$, direct computation gives $\pi_{-1} \fGP(\bZ_2^T) = \bZ_2$, whereas for the differential $\d'_\epsilon$, we have $\pi_{-1} \fGP(\bZ_2^T) = 0$.

These answers are appropriate. Without protecting by time reversal, the invertible $0+1d$ system corresponds to the odd complex line $\bC^{0|1} \in \SV_\bC^\times$. The two notions of time reversal correspond, respectively, to studying either real supervector spaces $\SV_\bR$ --- in which case there is an invertible odd object, namely $\bR^{0|1}$ --- or ``quaternionic'' supervector spaces ``$\SV_\bH$,'' which are supervector spaces in which the even part is real and the odd part is quaternionic. Since the underlying complex vector space of every $\bH$-module is even-dimensional, this latter category $\SV_\bH$ does not have an invertible odd object: $\pi_{-1} \SV_\bH^\times$ is the trivial group.

\subsection{Extended supercohomology with time reversal}

The two extensions of complex conjugation to $\SV_\bC$ provide two actions of $\bZ_2$ on $\fGP_{\leq 2} = \SA_\bC^\times$. In fact, there are four actions of $\bZ_2$ on $\SA_\bC^\times$ that induce the nontrivial action on $\pi_1 = \bZ$, two of which extend to the bicategory $\SA_\bC$ of all (not necessarily invertible) superalgebras and two of which do not. The actions that do not extend to all of $\SA_\bC$ don't seem to have a physical meaning: they are actions on the space $\GP_{\leq 2}$ of invertible phases $1+1d$ phases that do not make sense on noninvertible defects.

The ``$\d_\epsilon$'' action from \S\ref{gu wen time}, for which $\fGP(\bZ_2^T)_{\leq 1}$, gives an action on $\SA_\bC$ for which $\fGP(\bZ_2^T)_{\leq 2} = \SA_\bR^\times$ is the spectrum of Morita-invertible \define{real} superalgebras. The homotopy groups of this spectrum are well-known: up to Morita equivalence, there are eight real Clifford algebras, and so
$$ \pi_0 \fGP(\bZ_2^T)_{\leq 2} = \bZ_2, \quad \pi_{-1} \fGP(\bZ_2^T)_{\leq 2} = \bZ_2, \quad \pi_{-2} \fGP(\bZ_2^T)_{\leq 2} = \bZ_8.$$
It is not hard to show that both k-invariants are nontrivial, and an argument as in \S\ref{extended supercohomology} shows that the two choices of j-invariant determine the same spectrum.

Let us analyze the quaternionic (aka symplectic) case, which in \S\ref{gu wen time} provided the twisted differential $\d'_\epsilon$. For this action, the spectrum $\fGP(\bZ_2^T)_{\leq 2}$ can be modeled as the spectrum of Morita-invertible algebras in the category $\SV_\bH$ of quaternionic supervector spaces. $\SV_\bH$ is a somewhat unfamiliar category, and its algebras have a somewhat unfamiliar flavor. The irreducible odd object $I \in \SV_\bH$ is quaternionic in the sense that its endomorphism algebra is the quaternion algebra $\bH$, considered as a purely even algebra. In particular, $\bH$, being an endomorphism algebra, is Morita-trivial in this category. The nontrivial complex superalgebra $\Cliff_\bC(1) \in \SA_\bC^\times$ cannot be given a quaternionic structure. But $\Cliff_\bC(2)$, which is trivial in $\SA_\bC^\times$, does admit a nontrivial quaternionic structure, and so defines a nontrivial object in $\SA_\bH^\times$. The irreducible odd object $I \in \SV_\bH$ complexifies\footnote{The complexification of a real vector space $V$ is $V \otimes_\bR \bC$. The complexification of a quaternionic module is the underlying complex vector space.} to $\bC^{0|2}$, and the nontrivial object in $\SA_\bH^\times$ is ``$\Cliff(I)$.'' All together, we find:
$$ \pi_0 \fGP(\bZ_2^T)_{\leq 2} = \bZ_2, \quad \pi_{-1} \fGP(\bZ_2^T)_{\leq 2} = 0, \quad \pi_{-2} \fGP(\bZ_2^T)_{\leq 2} = \bZ_2.$$

Since there are only two nontrivial homotopy groups, the spectrum $\fGP(\bZ_2^T)_{\leq 2}$ will be determined once we can compute the k-invariant, a degree-$3$ stable cohomology operation. There are precisely four degree-$3$ stable cohomology operations from $\bZ_2$ to $\bZ_2$: $0$, $\Sq^1 \Sq^2$, $\Sq^2\Sq^1$, and $\Sq^1\Sq^2 + \Sq^2 \Sq^1$.

We first argue that the k-invariant is nontrivial. Consider the restricted $2+1d$ Gu--Wen phase for $G = \bZ_2$: it generates  $\H^3(\bZ_2; \fGP_{\leq 2}) = \bZ_4$, and so has order $4$. It can be realized as a monoidal $\bZ_2$-action on the Morita-trivial multifusion category of bimodules for the complex superalgebra $A = \bC \oplus \bC$. This monoidal category can be understood as the monoidal category $\operatorname{Mat}(2,\SV_\bC)$ of $2\times 2$ matrices whose matrix entries are complex supervector spaces. Denoting parity-reversal by $\Pi$, the $\bZ_2$-action is given by the functor $\bigl( \begin{smallmatrix} X & Y \\ Z & W \end{smallmatrix}\bigr) \mapsto \bigl( \begin{smallmatrix} W & \Pi Z \\ \Pi Y & X \end{smallmatrix} \bigr)$ plus monoidality data that we leave to the reader.\footnote{There are two consistent choices for this monoidality data, corresponding to the two order-$4$ elements in $\H^3(\bZ_2; \fGP_{\leq 2}) = \bZ_4$.}
We claim that this Gu--Wen phase can be given an auxiliary time-reversal structure with symplectic fermions. In fact, this can be done in multiple ways. One option is to use the category of bimodules for the quaternionic superalgebra $\bR \oplus \Cliff(I)$; such bimodules can again be realized as matrices $\bigl( \begin{smallmatrix} X & Y \\ Z & W \end{smallmatrix}\bigr)$ where now $X$ and $W$ are objects of $\SV_\bH$ and $Y$ and $Z$ are $\Cliff(I)$-modules. (Although ``parity reversal'' does not make sense in $\SV_\bH$, it does make sense for $\Cliff(I)$-modules, so the same formula $\bigl( \begin{smallmatrix} X & Y \\ Z & W \end{smallmatrix}\bigr) \mapsto \bigl( \begin{smallmatrix} W & \Pi Z \\ \Pi Y & X \end{smallmatrix} \bigr)$ still makes sense.) Another option is to use $\bC$ as a real algebra (hence as a quaternionic superalgebra), and to use a $\bZ_2$-action that mixes complex conjugation and parity-reversal. Either choice determines a class in $\H^3(\bZ_2; \fGP(\bZ_2^T)_{\leq 2})$ whose image in $\H^3(\bZ_2; \fGP)$ has order $4$. Thus $\H^3(\bZ_2; \fGP(\bZ_2^T)_{\leq 2})$ has elements of order at least~$4$, which is possible only when the k-invariant is nontrivial.

More generally, consider the map $\fGP(\bZ_2^T)_{\leq 2} \to \fGP_{\leq 2}$ that forgets the time-reversal symmetry (aka the complexification map $\SA_\bH^\times \to \SA_\bC^\times$). On homotopy groups this is the $0$ map: the nontrivial element $\Cliff(I) \in \pi_{-2} \fGP(\bZ_2^T)_{\leq 2} = \bZ_2$ maps to $\Cliff(2,\bC) \simeq \bC \in \pi_{-2} \fGP_{\leq 2}$.
$$ \begin{tikzpicture}[scale=1.5]
  \path
    (0,0) node (A) {$\fGP(\bZ_2^T)_{\leq 2}$}
    (3,0) node (A0) {$\bZ_2$}
    (4,0) node (A1) {$\phantom0$}
    (5,0) node (A2) {$\bZ_2$}
    (0,-1) node (B) {$\fGP_{\leq 2}$}
    (2,-1) node (Br) {$\bZ$}
    (3,-1) node (B0) {$\phantom0$}
    (4,-1) node (B1) {$\bZ_2$}
    (5,-1) node (B2) {$\bZ_2$}
  ;
  \draw[->] (A0) -- node[auto] {$\scriptstyle 0$} (B0);
  \draw[->] (A1) -- node[auto] {$\scriptstyle 0$} (B1);
  \draw[->] (A2) -- node[auto] {$\scriptstyle 0$} (B2);
\end{tikzpicture}$$
But complexification determines some nontrivial degree-$1$ maps. Indeed, on $\pi_0$ we have the inclusion $\bZ_2 \hookrightarrow \bC^\times$, which turns into the integral Bockstein $\Box : \bZ_2 \to \bZ$. The existence of a quaternionic structure on the Gu--Wen phase from the previous paragraph means that complexification induces the nontrivial map $\Sq^1 : \pi_{-2} \fGP(\bZ_2^T)_{\leq 2} = \bZ_2 \to \pi_{-1} \fGP_{\leq 2} = \bZ_2$. Letting $k$ denote the not-yet-determined k-invariant for $\fGP(\bZ_2^T)_{\leq 2}$, we have a commuting square:
$$ \begin{tikzpicture}[scale=1.5]
  \path
    (0,0) node (A) {$\fGP(\bZ_2^T)_{\leq 2}$}
    (3,0) node (A0) {$\bZ_2$}
    (4,0) node (A1) {}
    (5,0) node (A2) {$\bZ_2$}
    (0,-1) node (B) {$\fGP_{\leq 2}$}
    (2,-1) node (Br) {$\bZ$}
    (3,-1) node (B0) {}
    (4,-1) node (B1) {$\bZ_2$}
    (5,-1) node (B2) {$\bZ_2$}
  ;
  \draw[->] (A0) -- node[auto,swap] {$\scriptstyle \Box$} (Br);
  \draw[->] (A2) -- node[auto,swap] {$\scriptstyle \Sq^1$} (B1);
  \draw[->] (A2) .. controls +(-.5,.5) and +(.5,.5) .. node[auto,swap] {$\scriptstyle k$} (A0);
  \draw[->] (B1) .. controls +(-.5,-.5) and +(.5,-.5) .. node[auto,swap] {$\scriptstyle \Box\Sq^2$} (Br);
\end{tikzpicture}$$
and so
$$ \Box k = \Box \Sq^2 \Sq^1.$$
But $\Box\Sq^1 = 0$, leaving us with two choices for the k-invariant for $\fGP(\bZ_2^T)$: either $k = \Sq^2 \Sq^1$ or $k = \Sq^1 \Sq^2 + \Sq^2 \Sq^1$.

Further computations will be needed to determine which of these two options is the correct one to describe the spectrum $\fGP(\bZ_2^T)$ for symplectic fermions. Indeed, further computations are needed even to find the correct formula, akin to the matrices $\bigl( \begin{smallmatrix} -1 & 0 \\ & 1\end{smallmatrix}\bigr)$ and $\bigl( \begin{smallmatrix} -1 & \Box\Sq^1 \\ & 1\end{smallmatrix}\bigr)$ and the differentials $\d_\epsilon$ and $\d'_\epsilon$ from \S\ref{gu wen time}, for the two physically meaningful ``time-reversal'' actions of $\bZ_2$ on $\fGP_{\leq 2} = \SA_\bC^\times$ --- analyses that do not include noninvertible defects will not be able to rule out the nonphysical actions of $\bZ_2$ on $\SA_\bC^\times$ that do not extend to all of $\SA_\bC$.

\section{SPT phases as anomaly theories} \label{sec:anomalies}
SPT phases are often characterized by properties of their symmetry-preserving boundaries:  
the boundaries should support phases where the bulk symmetry is realized in ways which are impossible for 
a stand-alone system. 
Conversely, one may wonder how to define the notion of ``system with anomalous symmetry'' in such a way that the 
corresponding ``anomaly'' can be cured by an SPT phase in one dimension higher and ask if such anomaly fully 
captures the data of the corresponding bulk SPT phase. 

If we follow the idea that a non-anomalous internal symmetry is encoded in a collection of invertible gapped defects 
and junctions which satisfy appropriate associativity conditions, then an anomalous symmetry should be associated to 
collections of defects and junctions which fail to satisfy these associativity conditions. 

There are many ways these conditions may fail. For example, suppose we are given a phase equipped with 
a collection of invertible interfaces $U_g$, fusing according to a group law $U_g \circ U_{g'} \simeq U_{g g'}$. 
The next step in defining a non-anomalous symmetry would be to identify a viable set of invertible junctions 
$U_{g,g'}$ between these domain walls. 

The existence of such junctions is implicit in the statement $U_g \circ U_{g'} \simeq U_{g g'}$. The choice, though, 
is not unique. Different choices can be related by fusing the junctions with some invertible codimension $2$ defects 
$\lambda_2(g,g')$ in the bulk theory. Once we make some choice, we need to test associativity. Failure of associativity 
is also encoded in an invertible codimension $2$ defect $\alpha_3(g,g',g'')$. 

The candidate symmetry will be non-anomalous only if we can adjust the $\lambda_2(g,g')$ in such a way to cancel 
$\alpha_3(g,g',g'')$. This is a cohomology problem of sorts, but takes place in the world of invertible codimension $2$ defects
in the theory. This is not very good. Such a severe anomaly cannot be compared between different theories, nor cancelled by stacking 
the theory with some other theory in the same dimension or placing it at the boundary of an SPT in higher dimension. 

The situation is drastically ameliorated if the failure of associativity takes place in a smaller class of codimension $2$ defects
(possibly after adjusting our choice of junctions): those defined by stacking an invertible codimension $2$ phase onto the system. Then $\alpha_3(g,g',g'')$
is clearly a closed chain valued in $\GP_{d-1}$ and the anomaly is controlled by a group cohomology class valued in 
invertible phases in codimension $2$.

Similar considerations apply to higher associativity relations. The higher failures of associativity will be encoded in 
invertible defects of codimension $3$, $4$, etc. If these defects are generic, the ``anomaly'' of the symmetry 
cannot be compared between different theories nor cured by an higher-dimensional SPT phase. We are not interested in such a situation. 
But if the failures of associativity can be expressed in terms of ``exogenous'' defects, defined by staking 
the theory with a lower-dimensional invertible phase, then we have a good anomaly. 
The 3-dimensional failures of associativity will clearly build up a class in 
$\H^3(BG_b; \pi_0 \GP_{d-1})$. Along with higher-dimensional failures of associativity, the full data of the anomaly becomes a class in $\H^{d+2}(BG_b; \GP_{\leq d-1})$, which can be identified with an SPT phase in one dimension higher!

 With some hindsight, we can relax our assumptions a bit in order to accomodate the most general SPT phases 
 we have discussed. The first step of the relaxation is to accept that the symmetry interfaces may fuse 
 up to stacking with invertible phases in one dimension lower: 
 \begin{equation*}
 U_g \circ U_{g'} \simeq U_{g g'} \times \alpha_2(g,g')
 \end{equation*}
 This gives anomalies valued in $\H^2(BG_b;\pi_0\GP_{d})$.
 Finally, we may allow the interfaces themselves to be slightly ill-defined, in the sense that 
 $U_g$ is really an interface between the theory and the theory stacked with some invertible phase $\alpha_1(g)$. 
 This gives anomalies valued in $\H^1(BG_b;\pi_0\GP_{ d+1})$. All together these anomalies compile into an anomaly valued in $\H^{d+2}(BG_b; \GP_{\leq d+1}) = \tilde\H^{d+2}(BG_b; \GP)$.
 
We have thus identified a natural setup to define symmetries with a ``good'' anomaly, which can be compared between different 
theories and cured by a higher-dimensional SPT phase. Conversely, we have identified how the data of an SPT phase is encoded 
in the properties of a symmetry-preserving boundary system. 


\section*{Acknowledgements}
We thank Lakshya Bhardwaj and Apoorv Tiwari for useful conversations. We thank Dan Freed, Anton Kapustin and Edward Witten 
for important comments on the draft. 
The research of the authors was supported by the Perimeter Institute for Theoretical Physics. Research
at Perimeter Institute is supported by the Government of Canada through Industry
Canada and by the Province of Ontario through the Ministry of Economic Development and
Innovation.

\bibliographystyle{alpha}
\bibliography{lattice}

\end{document}